\documentstyle[12pt]{article}

\begin{document}

\title{Relativistic Quantum Coulomb Law}

\author{Yury M. Zinoviev\thanks{This work was supported in part by the Russian Foundation
 for Basic Research (Grant No. 07 - 01 - 00144) and Scientific Schools 672.2006.1.}}

\date{}
\maketitle

Steklov Mathematical Institute, Gubkin St. 8, 119991, Moscow, Russia,

 e - mail: zinoviev@mi.ras.ru

\vskip 1cm

\noindent {\bf Abstract.} The relativistic quantum mechanics
equations for the electromagnetic interaction are proposed.

\vskip 1cm

\section{Introduction}
\setcounter{equation}{0}

The celestial mechanics is based on the gravity law discovered by
Newton (1687). Cavendish (1773) proved by experiment that the force
of interaction between the electric charged bodies is inversely
proportional to the square of distance. This discovery was left
unpublished and later was repeated by de Coulomb (1785). The
electrodynamics equations were formulated by Maxwell (1873).  The
analysis of these equations led Lorentz (1904), Poincar\'e (1905,
1906), Einstein (1905) and Minkowski (1908) to the creation of the
theory of relativity.

Due to the paper \cite{1} the Maxwell equations are completely
defined by the relativistic Coulomb law. The relativistic Coulomb
law equations for two charged bodies are solved in the paper
\cite{2} for the case when one body moves freely. The results
\cite{2} may be applied to the study of the hydrogen atom. The light
electron moves around the heavy proton. The heavy proton moves
freely. The hydrogen spectrum is discrete and does not correspond to
the results of the paper \cite{2}. In order to study this problem we
need to apply the quantum electrodynamics. The following citation
from the book (\cite{4}, Chapter 4) describes the situation in the
quantum electrodynamics:

"But there is one additional problem that is characteristic of the
theory of quantum electrodynamics itself, which took twenty years to
overcome. It has to do with ideal electrons and photons and the
numbers $n$ and $j$.

"If electrons were ideal, and went from point to point in space -
time {\it only} by the direct path, then there would be no problem:
$n$ would be the mass of an electron (which we can determine by
observation), and $j$ would simply be its "charge" (the amplitude
for the electron to couple with a photon). It can also be determined
by experiment.

"But no such ideal electron exist. The mass we observed in the
laboratory is that of a {\it real} electron, which emits and absorbs
its own photons from time to time, and therefore depends on the
amplitude for coupling, $j$. And the charge we observe is between a
{\it real} electron and a {\it real} photon - which can form an
electron - positron pair from time to time - and therefore depends
on E(A to B), which involves $n$. Since the mass and charge of an
electron are affected by these and all other alternatives, the
experimentally measured mass, $m$, and experimentally measured
charge, $e$, of the electron are different from the numbers we use
in our calculations, $n$ and $j$.

"If there were a definite mathematical connection between $n$ and
$j$ on the one hand, and $m$ and $e$ on the other, there would still
be no problem: we would simply calculate what values of $n$ and $j$
we need to start with in order to end up with the observed values,
$m$ and $e$. (If our calculations didn't agree with $m$ and $e$, we
would jiggle the original $n$ and $j$ around until they did.)

"Let's see how we actually calculate $m$. We write a series of terms
that is something like the series we saw for the magnetic moment of
the electron: the first term has no couplings - just E(A to B) - and
represents an ideal electron going directly from point to point in
space - time. The second term has two couplings and represents a
photon being emitted and absorbed. Then come terms with four, six,
and eight couplings, and so on.

"When calculating terms with couplings, we must consider (as always)
all the possible points where couplings can occur, right down to
cases where the two coupling points are on top of each other - with
zero distance between them. The problem is, when we try to calculate
all the way down to zero distance, the equation blows up in our face
and gives meaningless answers - things like infinity. This caused a
lot of trouble when the theory of quantum electrodynamics first came
out. People were getting infinity for every problem they tried to
calculate! (One should be able to go down to zero distance in order
to be mathematically consistent, but that's there is no $n$ or $j$
that makes any sense; that's where the trouble is.)

"Well, instead of including all possible coupling points down to a
distance of zero, if one {\it stops} the calculation when the
distance between coupling points is very small - say, $10^{- 30}$
centimeters, billions and billions of times smaller than anything
observable in experiment (presently $10^{- 16}$ centimeters) - then
there are definite values for $n$ and $j$ that we can use so that
the calculated mass comes our to match the $m$ observed in
experiments, and the calculated charge matches the observed charge,
$e$. Now, here's the catch: if somebody else comes and stops their
calculation at a different distance - say, $10^{- 40}$ centimeters -
{\it their} values for $n$ and $j$ needed to get the same $m$ and
$e$ come out {\it different}!

"Twenty years later, in 1949, Hans Bethe and Victor Weisskopf
noticed something: if two people who stopped at different distances
to determine $n$ and $j$ from the same $m$ and $e$ then calculated
the answer to some {\it other} problem - each using the appropriate
but different values for $n$ and $j$ - when all the arrows from all
the terms were included, their answers to this other problem came
out nearly the same! In fact, the closer to zero distance that the
calculations for $n$ and $j$ were stopped, the better the final
answers for the other problem would agree! Schwinger, Tomonaga, and
I independently invented ways to make definite calculations to
confirm that it is true (we got prizes for that). People could
finally calculate with the theory of quantum electrodynamics!

"So it appears that the {\it only} thing that depend on the small
distance between coupling points are the values for $n$ and $j$ -
{\it theoretical numbers that are not directly observable anyway};
everything else, which {\it can} be observed, seems not to be
affected.

"The shell game that we play to find $n$ and $j$ is technically
called "renormalization". But no matter how clever the word, it is
what I would call a dippy process! Having to resort to such hocus -
pocus has prevented us from proving that the theory of quantum
electrodynamics is mathematically self - consistent. It's surprising
that the theory still hasn't been proved self - consistent one way
or the other by now; I suspect that renormalization is not
mathematically legitimate. What {\it is} certain is that we do not
have a good mathematical way to describe the theory of quantum
electrodynamics: such a bunch of words to describe the connection
between $n$ and $j$ and $m$ and $e$ is not good mathematics."

This paper is devoted to the relativistic quantum mechanics. The
equations for the electromagnetic interaction are proposed.

\section{Lorentz group}
\setcounter{equation}{0}

The theory of relativity has the mathematical foundation. It is
possible to add and multiply the complex numbers. Let us consider
the complex $2\times 2$ - matrices
\begin{equation}
\label{1.1} A =  \left( \begin{array}{cc}

A_{11} & A_{12} \\

A_{21} & A_{22}

\end{array} \right).
\end{equation}
The $2\times 2$ - matrix
\begin{equation}
\label{1.2} A^{\ast} = \left( \begin{array}{cc}

\bar{A}_{11} & \bar{A}_{21} \\

\bar{A}_{12} & \bar{A}_{22}

\end{array} \right)
\end{equation}
is called Hermitian adjoint matrix. If $A^{\ast} = A$, then the
matrix (\ref{1.1}) is Hermitian. Let us consider the basis of
Hermitian $2\times 2$ - matrices
\begin{equation}
\label{1.3} \sigma^{0} = \left( \begin{array}{cc}

1 & 0 \\

0 & 1

\end{array} \right),
\sigma^{1} = \left( \begin{array}{cc}

0 & 1 \\

1 & 0

\end{array} \right),
\sigma^{2} = \left( \begin{array}{cc}

0 & - i \\

i &   0

\end{array} \right),
\sigma^{3} = \left( \begin{array}{cc}

1 & 0 \\

0 & - 1

\end{array} \right).
\end{equation}
Any matrix (\ref{1.1}) has the form
\begin{equation}
\label{1.4} \sum_{\mu = 0}^{3} z^{\mu} \sigma^{\mu}
\end{equation}
where $z^{0}$,...,$z^{3}$ are the complex numbers. It is possible to
add and multiply the matrices (\ref{1.1}). These matrices form the
eight dimensional space of the matrices (\ref{1.4}) which is an
algebra. The following multiplication rules are valid
\begin{eqnarray}
\label{1.5} \sigma^{\mu} \sigma^{\mu} = \sigma^{0},\, \, \sigma^{0}
\sigma^{\mu} = \sigma^{\mu} \sigma^{0} = \sigma^{\mu}, \, \, \mu =
0,...,3; \nonumber \\
\sigma^{k_{1}} \sigma^{k_{2}} = \sum_{k_{3}\, =\, 1}^{3}
\epsilon^{k_{1}k_{2}k_{3}} i \sigma^{k_{3}}, \, \, k_{1},k_{2} =
1,2,3, \, \, k_{1} \neq k_{2}
\end{eqnarray}
where the antisymmetric tensor $\epsilon^{k_{1}k_{2}k_{3}}$ has the
normalization $\epsilon^{123} = 1$. Hence for the real numbers
$x^{0}$,...,$x^{3}$ the matrices
\begin{equation}
\label{1.6} \tilde{\tilde{x}} = x^{0}\sigma^{0} + i\sum_{k\, =\,
1}^{3} x^{k}\sigma^{k}
\end{equation}
form an algebra. The matrix (\ref{1.6}) is called quaternion. The
quaternion algebra was invented by Hamilton (1843). The algebra of
the matrices (\ref{1.6}) is the non - commutative extension of the
complex numbers field. The real numbers are the real diagonal
$2\times 2$ - matrices $x^{0}\sigma^{0}$. The pure imaginary numbers
are the real antisymmetric $2\times 2$ - matrices
$ix^{2}\sigma^{2}$. The determinant of a quaternion (\ref{1.6})
\begin{equation}
\label{1.7} \det \tilde{\tilde{x}} = \sum_{\mu \, =\, 0}^{3}
(x^{\mu})^{2}
\end{equation}
is called the Euclidean metric. Any matrix (\ref{1.1}) satisfying
the equation
\begin{equation}
\label{1.8} A^{\ast} = (\det A)A^{- 1}
\end{equation}
has the form (\ref{1.6}) and is a quaternion. The equation
(\ref{1.8}) implies also that the matrices (\ref{1.6}) form an
algebra. The matrices (\ref{1.6}) with determinant equal to $1$
satisfy the equations $A^{\ast}A = \sigma^{0}$, $\det A = 1$ and
form the group $SU(2)$. The matrices (\ref{1.1}) with determinant
equal to $1$ form the group $SL(2,{\bf C})$. The group $SU(2)$ is
the maximal compact subgroup of the group $SL(2,{\bf C})$. We
identify a vector $x^{\mu}$, $\mu = 0,...,3$, from the four
dimensional Euclidean space and a quaternion (\ref{1.6}). The unit
sphere in the four dimensional Euclidean space is isomorphic to the
group $SU(2)$. A rotation of the four dimensional Euclidean space is
a matrix product
\begin{equation}
\label{1.9} R(A,B)(\tilde{\tilde{x}}) = A\tilde{\tilde{x}} B
\end{equation}
where the matrices $A,B \in SU(2)$. The metric (\ref{1.7}) is not
changed under any rotation (\ref{1.9}).

In the quaternion (\ref{1.6}) the coordinate $x^{0}$ is a real
number and the coordinates $ix^{k}$, $k = 1,2,3$, are the pure
imaginary numbers. Let us consider the four dimensional space of
Hermitian matrices
\begin{equation}
\label{1.10} \tilde{x} = \sum_{\mu = 0}^{3} x^{\mu} \sigma^{\mu}
\end{equation}
where the coordinates $x^{0}$,...,$x^{3}$ are the real numbers. It
is possible to add $\tilde{x} + \tilde{y}$ and to multiply $1/2
(\tilde{x} \tilde{y} + \tilde{y} \tilde{x})$ these matrices. The
determinant of a matrix (\ref{1.10}) is
\begin{equation}
\label{1.11} \det \tilde{x} = (x^{0})^{2} - |{\bf x}|^{2} = (x,x) =
\sum_{\mu, \nu \, =\, 0}^{3} \eta_{\mu \nu} x^{\mu}x^{\nu}.
\end{equation}
Here the diagonal matrix $\eta_{\mu \nu} = \eta^{\mu \nu}$,
$\eta_{00} = - \, \eta_{11} = - \, \eta_{22} = - \, \eta_{33} = 1$.
The Minkowski metric (\ref{1.11}) was introduced by Poincar\'e
(1906). For the matrices $A,B \in SL(2,{\bf C})$ we consider a
transformation
\begin{equation}
\label{1.12} L(A,B)(\tilde{x}) = A\tilde{x} B
\end{equation}
similar to a rotation (\ref{1.9}). A transformation (\ref{1.12})
does not change a determinant (\ref{1.11}). A matrix (\ref{1.12}) is
Hermitian if
\begin{equation}
\label{1.13} B^{\ast}\tilde{x} A^{\ast} = A\tilde{x} B.
\end{equation}
A matrix $A$ is invertible since its determinant is equal to $1$.
Hence the relation (\ref{1.13}) implies
\begin{equation}
\label{1.14} A^{- 1}B^{\ast}\tilde{x} = \tilde{x} B(A^{\ast})^{- 1}.
\end{equation}
By inserting the matrix $\tilde{x} = \sigma^{0}$ into the equality
(\ref{1.14}) we have $A^{- 1}B^{\ast} = B(A^{\ast})^{- 1}$. The
Hermitian matrix $B(A^{\ast})^{- 1}$ commutes with any Hermitian
matrix. Hence
\begin{equation}
\label{1.15} B(A^{\ast})^{- 1} = \lambda \sigma^{0}
\end{equation}
where $\lambda$ is a real number. The number $\lambda = \pm 1$ since
the determinants of the matrices $A,B$ are equal to $1$. It is easy
to verify for any matrix (\ref{1.1})
\begin{equation}
\label{1.16} \hbox{tr} (L(A,\lambda A^{\ast})(\sigma^{0})) = \lambda
\sum_{i,j\, =\, 1}^{2} |A_{ij}|^{2}.
\end{equation}
The number (\ref{1.16}) is the double coefficient at the matrix
$\sigma^{0}$ in the decomposition (\ref{1.10}) for the matrix
$L(A,\lambda A^{\ast})(\sigma^{0})$. For $\lambda = 1$ the number
(\ref{1.16}) is positive and the time direction is not changed. The
group of the transformations $L(A,A^{\ast})$, $A \in SL(2,{\bf C})$,
is called the Lorentz group.

\section{Relativistic quantum laws}
\setcounter{equation}{0}

For a complex $2\times 2$ - matrix (\ref{1.1}) we define the
following $2\times 2$ - matrices
\begin{equation}
\label{2.1} A^{T} = \left( \begin{array}{cc}

A_{11} & A_{21} \\

A_{12} & A_{22}

\end{array} \right), \, \,
\bar{A} = \left( \begin{array}{cc}

\bar{A}_{11} & \bar{A}_{12} \\

\bar{A}_{21} & \bar{A}_{22}

\end{array} \right).
\end{equation}
Let us describe the irreducible representations of the group
$SU(2)$. We consider the half - integers $l \in 1/2{\bf Z}_{+}$,
i.e. $l = 0,1/2,1,3/2,...$. We define the representation of the
group $SU(2)$ on the space of the polynomials with degree less than
or equal to $2l$
\begin{equation}
\label{2.2} T_{l}(A)\phi (z) = (A_{12}z + A_{22})^{2l}\phi \left(
 \frac{A_{11}z + A_{21}}{A_{12}z + A_{22}}\right).
\end{equation}
We consider a half - integer $n = - l, - l + 1,...,l - 1,l$ and
choose the polynomial basis
\begin{equation}
\label{2.3} \psi_{n} (z) = ((l - n)!(l + n)!)^{- 1/2}z^{l - n}.
\end{equation}
The definitions (\ref{2.2}), (\ref{2.3}) imply
\begin{equation}
\label{2.4} T_{l}(A)\psi_{n} (z) = \sum_{m = - l}^{l} \psi_{m}
(z)t_{mn}^{l}(A),
\end{equation}
\begin{eqnarray}
\label{2.5} t_{mn}^{l}(A) = ((l - m)!(l + m)!(l - n)!(l + n)!)^{1/2}
\times \nonumber \\ \sum_{j = - \infty}^{\infty} \frac{A_{11}^{l - m
- j}A_{12}^{j}A_{21}^{m - n + j}A_{22}^{l + n - j}}{\Gamma (j +
1)\Gamma (l - m - j + 1)\Gamma (m - n + j + 1)\Gamma (l + n - j +
1)}
\end{eqnarray}
where $\Gamma (z)$ is the gamma - function. The function $(\Gamma
(z))^{- 1}$ equals zero for $z = 0 , - 1, - 2,...$. Therefore the
series (\ref{2.5}) is a polynomial.

The relation (\ref{2.2}) defines a representation of the group
$SU(2)$. Thus the polynomial (\ref{2.5}) defines a representation of
the group $SU(2)$
\begin{equation}
\label{2.6} t_{mn}^{l}(AB) = \sum_{k = - l}^{l}
t_{mk}^{l}(A)t_{kn}^{l}(B).
\end{equation}
This $(2l + 1)$ - dimensional representation is irreducible
(\cite{5}, Chapter III, Section 2.3). The relations (\ref{2.5}),
(\ref{2.6}) have an analytic continuation to all matrices
(\ref{1.1}).

By making the change $j \rightarrow j + n - m$ of the summation
variable in the equality (\ref{2.5}) we have
\begin{equation}
\label{2.7} t_{mn}^{l}(A) = t_{nm}^{l}(A^{T}).
\end{equation}

The polynomial (\ref{2.5}) is homogeneous of the matrix elements
(\ref{1.1}). Its degree is $2l$. The sum (\ref{1.5}) contains the
only non - zero term. Since the definition (\ref{2.5}) implies
\begin{equation}
\label{2.10} t_{mn}^{l}(\sigma^{0}) = \delta_{mn},
\end{equation}
the relations (\ref{1.5}), (\ref{2.6}) imply
\begin{eqnarray}
\label{2.8} \sum_{p\, =\, - l}^{l} \sum_{\dot{p} \, =\, -
\dot{l}}^{\dot{l}} \left( \sum_{\nu \, =\, 0}^{3}
t_{mp}^{l}(\sigma^{\nu})t_{\dot{m}
\dot{p}}^{\dot{l}}(\overline{\sigma^{\nu}}) \left( -
i\frac{\partial}{\partial x^{\nu}}\right) \right) \left( \sum_{\nu
\, =\, 0}^{3} \eta^{\nu \nu} t_{pn}^{l}(\sigma^{\nu})t_{\dot{p}
\dot{n}}^{\dot{l}}(\overline{\sigma^{\nu}})
\left( - i\frac{\partial}{\partial x^{\nu}} \right) \right) = \nonumber \\
\sum_{1\, \leq \, k_{1}\, <\, k_{2}\, \leq \, 3} \sum_{k_{3}\, =\,
1}^{3} t_{mn}^{l}(i\sigma^{k_{3}})t_{\dot{m} \dot{n}}^{\dot{l}}(-
i\overline{\sigma^{k_{3}}})((\epsilon^{k_{1}k_{2}k_{3}})^{2l +
2\dot{l}} + (\epsilon^{k_{2}k_{1}k_{3}})^{2l +
2\dot{l}})\frac{\partial^{2}}{\partial x^{k_{1}}\partial
x^{k_{2}}}\nonumber \\ - \delta_{mn} \delta_{\dot{m} \dot{n}}
(\partial_{x}, \partial_{x}),\, \, (\partial_{x},
\partial_{x}) = \sum_{\nu \, =\, 0}^{3} \eta^{\nu \nu} \left(
\frac{\partial}{\partial x^{\nu}} \right)^{2}.
\end{eqnarray}
For an odd integer $2l + 2\dot{l}$ the relation (\ref{2.8}) has the
form
\begin{eqnarray}
\label{2.9} \sum_{p\, =\, - l}^{l} \sum_{\dot{p} \, =\, -
\dot{l}}^{\dot{l}} \left( \sum_{\nu \, =\, 0}^{3}
t_{mp}^{l}(\sigma^{\nu})t_{\dot{m}
\dot{p}}^{\dot{l}}(\overline{\sigma^{\nu}}) \left( -
i\frac{\partial}{\partial x^{\nu}}\right) \right) \left( \sum_{\nu
\, =\, 0}^{3} \eta^{\nu \nu} t_{pn}^{l}(\sigma^{\nu})t_{\dot{p}
\dot{n}}^{\dot{l}}(\overline{\sigma^{\nu}})\left( -
i\frac{\partial}{\partial x^{\nu}}\right) \right) = \nonumber \\  -
\delta_{mn} \delta_{\dot{m} \dot{n}} (\partial_{x},
\partial_{x}).
\end{eqnarray}
By making use of the relation (\ref{2.9}) for the odd integer $2l +
2\dot{l}$ the Lorentz invariant equation
\begin{equation}
\label{2.11} ( - (\partial_{x}, \partial_{x}) - \mu^{2})
\phi_{m\dot{m}} (x) = 0, \, \, m = - l, - l + 1,...,l - 1,l,\, \,
\dot{m} = - \dot{l}, - \dot{l} + 1,...,\dot{l} - 1,\dot{l}
\end{equation}
may be rewritten as the system of the linear equations
\begin{eqnarray}
\label{2.12} \sum_{n\, =\, 2l + 2}^{4l + 2} \sum_{\dot{n} \, =\,
1}^{2\dot{l} + 1} \sum_{\nu\, =\, 0}^{3} \eta^{\nu \nu} t_{m - l -
1,n - 3l - 2}^{l}(\sigma^{\nu}) t_{\dot{m} - \dot{l} - 1,\dot{n} -
\dot{l} - 1}^{\dot{l}}(\overline{\sigma^{\nu}})\left( -
i\frac{\partial}{\partial x^{\nu}} \right) \psi_{n\dot{n}} (x) +
\psi_{m\dot{m}} (x) = 0,\nonumber \\  m = 1,..., 2l + 1,\, \,
\dot{m} = 1,...,2\dot{l} + 1;\nonumber \\ \sum_{n\, =\, 1}^{2l + 1}
\sum_{\dot{n} \, =\, 1}^{2\dot{l} + 1} \sum_{\nu\, =\, 0}^{3} t_{m -
3l - 2,n - l - 1}^{l}(\sigma^{\nu}) t_{\dot{m} - \dot{l} - 1,\dot{n}
- \dot{l} - 1}^{\dot{l}}(\overline{\sigma^{\nu}})\left( -
i\frac{\partial}{\partial x^{\nu}} \right) \psi_{n\dot{n}} (x) +
\mu^{2} \psi_{m\dot{m}} (x) = 0,\nonumber \\  m = 2l + 2,..., 4l +
2,\, \, \dot{m} = 1,...,2\dot{l} + 1.
\end{eqnarray}
Let us define the $((4l + 2)(2\dot{l} + 1))\times ((4l + 2)(2\dot{l}
+ 1))$ - matrices
\begin{eqnarray}
\label{2.13} (\alpha_{l,\dot{l}} (\mu^{2}))_{m\dot{m},n\dot{n}} =
\mu^{2} \delta_{mn} \delta_{\dot{m} \dot{n}}, \, \, m,n = 1,...,2l +
1,\, \, \dot{m}, \dot{n} = 1,...,2\dot{l} + 1;\nonumber
\\ (\alpha_{l,\dot{l}} (\mu^{2}))_{m\dot{m},n\dot{n}} =
\delta_{mn} \delta_{\dot{m} \dot{n}}, \, \, m,n = 2l + 2,...,4l +
2,\, \, \dot{m}, \dot{n} = 1,...,2\dot{l} + 1; \nonumber \\
(\beta_{l,\dot{l}} (\mu^{2}))_{m\dot{m},n\dot{n}} = \delta_{mn}
\delta_{\dot{m} \dot{n}}, \, \, m,n = 1,...,2l +
1,\, \, \dot{m}, \dot{n} = 1,...,2\dot{l} + 1;\nonumber \\
(\beta_{l,\dot{l}} (\mu^{2}))_{m\dot{m},n\dot{n}} = \mu^{2}
\delta_{mn} \delta_{\dot{m} \dot{n}}, \, \, m,n = 2l + 2,...,4l +
2,\, \, \dot{m}, \dot{n} = 1,...,2\dot{l} + 1; \nonumber
\\ (\gamma_{l,\dot{l}}^{\nu} (\sigma^{0}))_{m\dot{m}, n\dot{n}} =
\eta^{\nu \nu} t_{m - l - 1,n - 3l - 2}^{l}(\sigma^{\nu}) t_{\dot{m}
- \dot{l} - 1,\dot{n} - \dot{l} -
1}^{\dot{l}}(\overline{\sigma^{\nu}}),\nonumber \\ m = 1,...,2l +
1,\, \, n = 2l + 2,...,4l + 2,\, \, \dot{m}, \dot{n} = 1,...,
2\dot{l} + 1,\, \, \nu = 0,...,3; \nonumber
\\ (\gamma_{l,\dot{l}}^{\nu} (\sigma^{0}))_{m\dot{m}, n\dot{n}} =
t_{m - 3l - 2,n - l - 1}^{l}(\sigma^{\nu}) t_{\dot{m} - \dot{l} -
1,\dot{n} - \dot{l} -
1}^{\dot{l}}(\overline{\sigma^{\nu}}),\nonumber
\\ m = 2l + 2,...,4l + 2,\, \, n = 1,...,2l + 1,\, \, \dot{m},
\dot{n} = 1,..., 2\dot{l} + 1,\, \, \nu = 0,...,3.
\end{eqnarray}
The other matrix elements are equal to zero. If an integer $2l +
2\dot{l}$ is odd, then an integer $(2l + 1)(2\dot{l} + 1)$ is even
and an integer $(4l + 2)(2\dot{l} + 1)$ has the form $4k$ where $k$
is an integer.

By making use of the definitions (\ref{2.13}) we can rewrite the
equations (\ref{2.12}) as
\begin{equation}
\label{2.14} \sum_{n\, =\, 1}^{4l + 2} \sum_{\dot{n} \, =\,
1}^{2\dot{l} + 1} \left( \sum_{\nu \, =\, 0}^{3}
(\gamma_{l,\dot{l}}^{\nu} (\sigma^{0}))_{m\dot{m}, n\dot{n}} \left(
- i\frac{\partial}{\partial x^{\nu}} \right) + (\beta_{l,\dot{l}}
(\mu^{2}))_{m\dot{m}, n\dot{n}} \right) \psi_{n\dot{n}} (x) = 0.
\end{equation}
The definitions (\ref{2.13}) imply
\begin{equation}
\label{2.15} \alpha_{l,\dot{l}} (\mu) \beta_{l,\dot{l}} (\mu^{2}) =
\mu  \beta_{l,\dot{l}} (\mu),\, \, \alpha_{l,\dot{l}} (\mu)
\gamma_{l,\dot{l}}^{\nu} (\sigma^{0}) = \gamma_{l,\dot{l}}^{\nu}
(\sigma^{0}) \beta_{l,\dot{l}} (\mu).
\end{equation}
In view of the relations (\ref{2.15})  the action of the matrix
$\alpha_{l,\dot{l}} (\mu)$ on the equation (\ref{2.14}) yields
\begin{eqnarray}
\label{2.16} \sum_{n\, =\, 1}^{4l + 2} \sum_{\dot{n} \, =\,
1}^{2\dot{l} + 1}  \sum_{\nu \, =\, 0}^{3} (\gamma_{l,\dot{l}}^{\nu}
(\sigma^{0}))_{m\dot{m}, n\dot{n}} \left( -
i\frac{\partial}{\partial x^{\nu}} \right) \xi_{n\dot{n}} (x) + \mu
\xi_{m\dot{m}} (x) = 0, \nonumber \\ \xi_{m\dot{m}} (x) = \sum_{n\,
=\, 1}^{4l + 2} \sum_{\dot{n} \, =\, 1}^{2\dot{l} + 1}
(\beta_{l,\dot{l}} (\mu))_{m\dot{m}, n\dot{n}} \psi_{n\dot{n}} (x) =
0.
\end{eqnarray}
For $\mu > 0$ the transformation given by the second relation
(\ref{2.16}) is the isomorphism. The definition (\ref{2.5}) implies
\begin{equation}
\label{2.17} t_{mn}^{\frac{1}{2}} (A) = A_{m + \frac{3}{2}, n +
\frac{3}{2}}.
\end{equation}
Due to the relations (\ref{2.13}), (\ref{2.17}) the equation
(\ref{2.16}) for $l = \frac{1}{2}$, $\dot{l} = 0$ coincides with the
Dirac equation (\cite{6}, equation (1 - 41)).

The relations (\ref{1.10}), (\ref{1.12}) define the representation
of the group $SL(2,{\bf C})$ in the Lorentz group
\begin{equation}
\label{2.18} \sum_{\mu, \nu \, =\, 0}^{3} \Lambda_{\nu}^{\mu} (A)
x^{\nu}\sigma^{\mu} = L(A,A^{\ast})(\tilde{x}) = A\tilde{x}
A^{\ast}.
\end{equation}
Let us define $(4l + 2)(2\dot{l} + 1)$ - dimensional representation
of the group $SL(2,{\bf C})$
\begin{eqnarray}
\label{2.19} (S_{l,\dot{l}}(A))_{m\dot{m}, n\dot{n}} = t_{m - l -
1,n - l - 1}^{l}(A)t_{\dot{m} - \dot{l} - 1,\dot{n} - \dot{l} -
1}^{\dot{l}}(\bar{A}), \nonumber \\  m,n = 1,...,2l + 1,\, \,
\dot{m}, \dot{n} = 1,...,2\dot{l} + 1; \nonumber
\\ (S_{l,\dot{l}}(A))_{m\dot{m}, n\dot{n}} = t_{m - 3l - 2,n - 3l -
2}^{l}((A^{\ast})^{- 1})t_{\dot{m} - \dot{l} - 1,\dot{n} - \dot{l} -
1}^{\dot{l}}((A^{T})^{- 1}), \nonumber \\ m,n = 2l + 2,...,4l + 2,\,
\, \dot{m}, \dot{n} = 1,...,2\dot{l} + 1.
\end{eqnarray}
The other matrix elements are equal to zero. The definitions
(\ref{2.13}), (\ref{2.19}) imply
\begin{equation}
\label{2.20} S_{l,\dot{l}}(A)\alpha_{l,\dot{l}}(\mu^{2})
S_{l,\dot{l}}(A^{- 1}) = \alpha_{l,\dot{l}} (\mu^{2}), \, \,
S_{l,\dot{l}}(A)\beta_{l,\dot{l}}(\mu^{2}) S_{l,\dot{l}}(A^{- 1}) =
\beta_{l,\dot{l}} (\mu^{2}).
\end{equation}
Let the functions $\psi_{m\dot{m}} (x)$, $m = 1,...,4l + 2$,
$\dot{m} = 1,...,2\dot{l} + 1$ be the solutions of the equation
(\ref{2.14}). The relations (\ref{2.20}) imply that the functions
\begin{equation}
\label{2.21} \xi_{m\dot{m}} (x) = \sum_{n\, =\, 1}^{4l + 2}
\sum_{\dot{n} \, =\, 1}^{2\dot{l} + 1} (S_{l,\dot{l}}(A))_{m\dot{m},
n\dot{n}} \psi_{n\dot{n}} \left( \sum_{\nu \, =\, 0}^{3}
\Lambda_{\nu}^{\mu} (A^{- 1})x^{\nu} \right)
\end{equation}
are the solutions of the equation
\begin{equation}
\label{2.22} \sum_{n\, =\, 1}^{4l + 2} \sum_{\dot{n} \, =\,
1}^{2\dot{l} + 1} \left( \sum_{\nu \, = \, 0}^{3}
(\gamma_{l,\dot{l}}^{\nu} (A))_{m\dot{m}, n\dot{n}} \left( -
i\frac{\partial}{\partial x^{\nu}}\right) + (\beta_{l,\dot{l}}
(\mu^{2}))_{m\dot{m}, n\dot{n}} \right) \xi_{n\dot{n}} (x) = 0,
\end{equation}
\begin{equation}
\label{2.23} \gamma_{l,\dot{l}}^{\mu} (A) = \sum_{\nu \, =\, 0}^{3}
\Lambda_{\nu}^{\mu} (A)S_{l,\dot{l}} (A)\gamma_{l,\dot{l}}^{\nu}
(\sigma^{0}) S_{l,\dot{l}} (A^{- 1})
\end{equation}
for any matrix $A \in SL(2,{\bf C})$. The definition (\ref{2.23})
implies
\begin{equation}
\label{2.24} \gamma_{l,\dot{l}}^{\mu} (AB) = \sum_{\nu \, =\, 0}^{3}
\Lambda_{\nu}^{\mu} (A)S_{l,\dot{l}} (A)\gamma_{l,\dot{l}}^{\nu} (B)
S_{l,\dot{l}} (A^{- 1})
\end{equation}
for any matrices $A,B \in SL(2,{\bf C})$. By changing the coordinate
system we change the matrix $\gamma_{l,\dot{l}}^{\nu} (\sigma^{0})$
in the equation (\ref{2.14}) for the matrix (\ref{2.23}). The
solutions of the equation (\ref{2.14}) transform to the solutions
(\ref{2.21}) of the equation (\ref{2.22}). It is valid for all half
- integers $l,\dot{l} \in 1/2 {\bf Z}_{+}$. Due to (\cite{6},
relation (1 - 43))
\begin{equation}
\label{2.25} \gamma_{\frac{1}{2}, 0}^{\mu} (A) =
\gamma_{\frac{1}{2}, 0}^{\mu} (\sigma^{0})
\end{equation}
for any matrix $A \in SL(2,{\bf C})$. Hence the equation
(\ref{2.14}) for $l = \frac{1}{2}$, $\dot{l} = 0$ is covariant under
the group $SL(2,{\bf C})$.

In view of the definitions (\ref{2.13}), (\ref{2.19}), (\ref{2.23})
the equation (\ref{2.22}) is equivalent to the system of two
equations
\begin{eqnarray}
\label{2.26} \sum_{n\, =\, 2l + 2}^{4l + 2} \sum_{\dot{n} \, =\,
1}^{2\dot{l} + 1} \sum_{\nu \, =\, 0}^{3} (\gamma_{l,\dot{l}}^{\nu}
(A))_{m\dot{m}, n\dot{n}} \left( - i\frac{\partial}{\partial
x^{\nu}} \right) \xi_{n\dot{n}} (x) + \xi_{m\dot{m}} (x) = 0,
\nonumber \\ m = 1,...,2l + 1,\, \, \dot{m} = 1,...,2\dot{l} + 1;
\nonumber \\ \sum_{n\, =\, 1}^{2l + 1} \sum_{\dot{n} \, =\,
1}^{2\dot{l} + 1} \sum_{\nu \, =\, 0}^{3} (\gamma_{l,\dot{l}}^{\nu}
(A))_{m\dot{m}, n\dot{n}} \left( - i\frac{\partial}{\partial
x^{\nu}} \right) \xi_{n\dot{n}} (x) + \mu^{2} \xi_{m\dot{m}} (x) =
0, \nonumber \\ m = 2l + 2,...,4l + 2,\, \, \dot{m} = 1,...,2\dot{l}
+ 1.
\end{eqnarray}
The system of two equations (\ref{2.26}) is equivalent to the
equation
\begin{eqnarray}
\label{2.27} \sum_{n\, =\, 2l + 2}^{4l + 2} \sum_{\dot{n} \, =\,
1}^{2\dot{l} + 1} \sum_{p\, =\, 1}^{2l + 1} \sum_{\dot{p} \, =\,
1}^{2\dot{l} + 1} \left( \sum_{\nu \, =\, 0}^{3}
(\gamma_{l,\dot{l}}^{\nu} (A))_{m\dot{m}, p\dot{p}}
\frac{\partial}{\partial x^{\nu}} \right) \left( \sum_{\nu \, =\,
0}^{3} (\gamma_{l,\dot{l}}^{\nu} (A))_{p\dot{p},
n\dot{n}}\frac{\partial}{\partial x^{\nu}} \right) \xi_{n\dot{n}}
(x) \nonumber \\ + \mu^{2} \xi_{m\dot{m}} (x) = 0, \, \, m = 2l +
2,...,4l + 2,\, \, \dot{m} = 1,...,2\dot{l} + 1.
\end{eqnarray}
Let us prove that for an odd integer $2l + 2\dot{l}$
\begin{eqnarray}
\label{2.28} \sum_{p\, =\, 1}^{2l + 1} \sum_{\dot{p} \, =\,
1}^{2\dot{l} + 1} \left( \sum_{\nu \, =\, 0}^{3}
(\gamma_{l,\dot{l}}^{\nu} (A))_{m\dot{m}, p\dot{p}}
\frac{\partial}{\partial x^{\nu}} \right) \left( \sum_{\nu \, =\,
0}^{3} (\gamma_{l,\dot{l}}^{\nu} (A))_{p\dot{p}, n\dot{n}}
\frac{\partial}{\partial x^{\nu}} \right) = (\partial_{x},
\partial_{x}) \delta_{mn} \delta_{\dot{m} \dot{n}}, \nonumber \\ m,n
= 2l + 2,...,4l + 2,\, \, \dot{m}, \dot{n} = 1,...,2\dot{l} + 1.
\end{eqnarray}
We denote
\begin{equation}
\label{2.29} \frac{\partial}{\partial y^{\mu}} = \sum_{\nu \, =\,
0}^{3} \Lambda_{\mu}^{\nu} (A)\frac{\partial}{\partial x^{\nu}}.
\end{equation}
The matrix $\Lambda_{\mu}^{\nu} (A)$ belongs to the Lorentz group
and the definition (\ref{2.29}) implies
\begin{equation}
\label{2.31} \sum_{\mu \, =\, 0}^{3} \eta^{\mu \mu} \left(
\frac{\partial}{\partial y^{\mu}} \right)^{2} = (\partial_{x},
\partial_{x}).
\end{equation}
The definitions (\ref{2.13}), (\ref{2.19}), (\ref{2.23}) and the
relations (\ref{2.9}), (\ref{2.31}) imply the equality (\ref{2.28}).
Therefore for an odd integer $2l + 2\dot{l}$ the equation
(\ref{2.22}) is equivalent to the equation (\ref{2.11}).

The relations (\ref{2.13}) imply
\begin{equation}
\label{2.401} (\alpha_{l,\dot{l}} (\mu^{2}) \beta_{l,\dot{l}}
(\mu^{2}))_{m\dot{m}, n\dot{n}} = \mu^{2} \delta_{mn}
\delta_{\dot{m} \dot{n}}, \, \, m,n = 1,...,4l + 2,\, \dot{m},
\dot{n} = 1,...,2\dot{l} + 1.
\end{equation}
In view of the second relation (\ref{2.15}) and the relations
(\ref{2.20}), (\ref{2.23}), (\ref{2.401}) the action of the matrix
$\gamma_{l,\dot{l}}^{0} (\sigma^{0}) \alpha_{l,\dot{l}} (\mu^{2})$
on the equation (\ref{2.14}) yields
\begin{equation}
\label{2.402} \sum_{n\, =\, 1}^{4l + 2} \sum_{\dot{n} \, =\,
1}^{2\dot{l} + 1} \{ \sum_{\nu \, =\, 0}^{3} (\gamma_{l,\dot{l}}^{0}
(\sigma^{0}) \gamma_{l,\dot{l}}^{\nu} (\sigma^{0}) \beta_{l,\dot{l}}
(\mu^{2}) )_{m\dot{m}, n\dot{n}} \left( - i\frac{\partial}{\partial
x^{\nu}}\right)  +  \mu^{2} (\gamma_{l,\dot{l}}^{0} (\sigma^{0})
)_{m\dot{m}, n\dot{n}} \} \psi_{n\dot{n}} (x) = 0.
\end{equation}
The relations (\ref{1.5}), (\ref{2.6}), (\ref{2.10}), (\ref{2.13})
imply
\begin{eqnarray}
\label{2.403} ((\gamma_{l,\dot{l}}^{0} (\sigma^{0}))^{2})_{m\dot{m},
n\dot{n}} = \delta_{mn} \delta_{\dot{m} \dot{n}}, \, \, m,n =
1,...,4l + 2,\, \, \dot{m}, \dot{n} = 1,...,2\dot{l} + 1; \nonumber
\\ (\gamma_{l,\dot{l}}^{0} (\sigma^{0}) \gamma_{l,\dot{l}}^{k}
(\sigma^{0}))_{m\dot{m}, n\dot{n}} = t_{m - l - 1,n - l - 1}^{l}
(\sigma^{k}) t_{m - \dot{l} - 1,n - \dot{l} - 1}^{\dot{l}}
(\overline{\sigma^{k}}), \nonumber \\ m,n = 1,...,2l + 1,\, \,
\dot{m}, \dot{n} = 1,...,2\dot{l} + 1,\, \, k = 1,2,3; \nonumber
\\ (\gamma_{l,\dot{l}}^{0} (\sigma^{0}) \gamma_{l,\dot{l}}^{k}
(\sigma^{0}))_{m\dot{m}, n\dot{n}} = - t_{m - 3l - 2,n - 3l - 2}^{l}
(\sigma^{k}) t_{m - \dot{l} - 1,n - \dot{l} - 1}^{\dot{l}}
(\overline{\sigma^{k}}), \nonumber \\ m,n = 2l + 2,...,4l + 2,\, \,
\dot{m}, \dot{n} = 1,...,2\dot{l} + 1,\, \, k = 1,2,3.
\end{eqnarray}
The other matrix elements are equal to zero. The coefficients of the
polynomial (\ref{2.5}) are real. Hence in view of the relations
(\ref{2.7}), (\ref{2.13}), (\ref{2.403}) the matrices
$\gamma_{l,\dot{l}}^{0} (\sigma^{0})$, $\gamma_{l,\dot{l}}^{0}
(\sigma^{0}) \gamma_{l,\dot{l}}^{\nu} (\sigma^{0}) \beta_{l,\dot{l}}
(\mu^{2})$, $\nu = 0,...,3$, are Hermitian. Due to the first
relation (\ref{2.403}) we have $(\gamma_{l,\dot{l}}^{0}
(\sigma^{0}))^{2} \beta_{l,\dot{l}} (\mu^{2}) = \beta_{l,\dot{l}}
(\mu^{2})$. Let the functions $\xi_{m\dot{m}} (x)$ have the form
(\ref{2.21}). Now the equation (\ref{2.402}) implies that the
integral
\begin{equation}
\label{2.404} \int d^{3}{\bf x} \sum_{m\, =\, 1}^{4l + 2}
\sum_{\dot{m} \, =\, 1}^{2\dot{l} + 1} (\beta_{l,\dot{l}}
(\mu^{2}))_{m\dot{m}, m\dot{m}} \Biggl| \sum_{n\, =\, 1}^{4l + 2}
\sum_{\dot{n} \, =\, 1}^{2\dot{l} + 1} (S_{l,\dot{l}} (A^{-
1}))_{m\dot{m}, n\dot{n}} \xi_{n\dot{n}} \left( \sum_{\nu \, =\,
0}^{3} \Lambda_{\nu}^{\lambda} (A)x^{\nu} \right) \Biggr|^{2}
\end{equation}
is independent of the variable $x^{0}$ for $x^{0} > 0$. The
integrand (\ref{2.404}) is called the probability density of a
solution of the equation (\ref{2.22}). For $A = \sigma^{0}$ the
integrand (\ref{2.404}) coincides with the usual probability density
for the function (\ref{2.16}). In the quantum mechanics the fixed
probability density defines Hilbert space where any Hamiltonian
acts. The integral (\ref{2.404}) depends on the parameter $\mu^{2}$
in the equation (\ref{2.22}). We do not expect the solutions of the
equations with interaction to have the independent of time integrals
similar to (\ref{2.404}). This mathematical assumption does not seem
physically reasonable. We deal with the asymptotic solutions of the
equations with interaction in an experiment. We expect the solutions
of the equations with interaction to coincide asymptotically with
the products of the solutions of the equation (\ref{2.22}). The
probability density of the last solutions is given by the integrand
(\ref{2.404}).

Let the functions $\xi_{m\dot{m}} (x)$ be the solutions of the
equation (\ref{2.22}). Let us introduce the distributions
\begin{equation}
\label{2.30} f_{m\dot{m}} (x) = \theta (x^{0})\xi_{m\dot{m}} (x),\,
\, \theta (x) = \left\{ {1, \hskip 0,5cm x \geq 0,} \atop {0, \hskip
0,5cm x < 0.} \right.
\end{equation}
The equation (\ref{2.22}) implies
\begin{eqnarray}
\label{2.32} \sum_{n\, =\, 1}^{4l + 2} \sum_{\dot{n} \, =\,
1}^{2\dot{l} + 1}  \left( \sum_{\nu \, = \, 0}^{3}
(\gamma_{l,\dot{l}}^{\nu} (A))_{m\dot{m}, n\dot{n}} \left( -
i\frac{\partial}{\partial x^{\nu}}\right) + (\beta_{l,\dot{l}}
(\mu^{2}))_{m\dot{m}, n\dot{n}} \right) f_{n\dot{n}} (x) = \nonumber
\\ - i \delta (x^{0})f^{0}_{m\dot{m}} (+ 0,{\bf x}),\, \, {\bf x} =
(x^{1},x^{2},x^{3}) \in {\bf R}^{3},
\end{eqnarray}
\begin{equation}
\label{2.66} f^{0}_{m\dot{m}} (+ 0,{\bf x}) = \sum_{n\, =\, 1}^{4l +
2} \sum_{\dot{n} \, =\, 1}^{2\dot{l} + 1} (\gamma_{l,\dot{l}}^{0}
(A))_{m\dot{m}, n\dot{n}} f_{n\dot{n}} (+ 0,{\bf x}).
\end{equation}

Let a support of a distribution $e_{\mu_{1}^{2},..., \mu_{n}^{2}}
(x) \in S^{\prime} ({\bf R}^{4})$ lie in the closed upper light
cone. Let a distribution $e_{\mu_{1}^{2},..., \mu_{n}^{2}} (x)$
satisfy the equation
\begin{equation}
\label{2.33} \left( \prod_{i\, =\, 1}^{n} (- (\partial_{x},
\partial_{x}) - \mu_{i}^{2}) \right) e_{\mu_{1}^{2},..., \mu_{n}^{2}} (x) =
\delta (x).
\end{equation}
By changing the differential operator $- (\partial_{x},
\partial_{x})$ for the differential operator
$$
\prod_{i\, =\, 1}^{n} (- (\partial_{x},
\partial_{x}) - \mu_{i}^{2})
$$
in the proof of Lemma 3 from the paper \cite{1} we obtain the
uniqueness of the distribution $e_{\mu_{1}^{2},..., \mu_{n}^{2}}
(x)$. Due to (\cite{3}, Section 30)
\begin{equation}
\label{2.34} e_{0}(x) = -\, (2\pi)^{- 1} \theta (x^{0})\delta
((x,x)),\, \, e_{0,0}(x) = (8\pi)^{- 1}\theta (x^{0})\theta ((x,x)).
\end{equation}
The second definition (\ref{2.30}) implies
\begin{equation}
\label{2.35} (\partial_{x}, \partial_{x}) (\theta (x^{0})\theta
((x,x))(x,x)^{n}) = 4n(n + 1)\theta (x^{0})\theta ((x,x))(x,x)^{n -
1},\, \, n = 1,2,....
\end{equation}
Due to the relations (\ref{2.34}), (\ref{2.35}) the distribution
$e_{0,...,0}(x)$ with $n$ zeros has the form
\begin{equation}
\label{2.36} e_{0,...,0}(x) = (- 1)^{n}(2\pi 4^{n - 1}(n - 2)!(n -
1)!)^{- 1}\theta (x^{0})\theta ((x,x))(x,x)^{n - 2},\, \, n =
2,3,....
\end{equation}
Let us prove
\begin{eqnarray}
\label{2.37} e_{\mu_{1}^{2},..., \mu_{n}^{2}} (x) = \lim_{\epsilon
\rightarrow + 0} (2\pi)^{- 4} \int d^{4}p\exp \{ - i(p,x)\}
\prod_{j\, =\, 1}^{n} ((p^{0} + i\epsilon )^{2} - |{\bf p}|^{2} -
\mu_{j}^{2} )^{- 1},\nonumber \\ (x,y) = x^{0}y^{0} - \sum_{k\, =\,
1}^{3} x^{k}y^{k}.
\end{eqnarray}
The integral (\ref{2.37}) is the solution of the equation
(\ref{2.33}). By making the shift of the integration path in the
right - hand side of the equality (\ref{2.37}) we obtain that the
distribution (\ref{2.37}) is equal to zero for $x^{0} < 0$. The
distribution (\ref{2.37}) is Lorentz invariant. Hence its support
lies in the closed upper light cone. Now the uniqueness of the
distribution (\ref{2.37}) implies the equality (\ref{2.37}).

For an odd integer $2l + 2\dot{l}$ the relation
\begin{eqnarray}
\label{2.38} \sum_{p\, =\, 1}^{4l + 2} \sum_{\dot{p} \, =\,
1}^{2\dot{l} + 1} \left( \sum_{\nu \, =\, 0}^{3}
(\gamma_{l,\dot{l}}^{\nu} (A))_{m\dot{m}, p\dot{p}} \left( -
i\frac{\partial}{\partial x^{\nu}} \right) + (\beta_{l,\dot{l}}
(\mu^{2}))_{m\dot{m}, p\dot{p}}\right) \times \nonumber
\\ \left( \sum_{\nu \, =\, 0}^{3} (\gamma_{l,\dot{l}}^{\nu}
(A))_{p\dot{p}, n\dot{n}} \left( - i\frac{\partial}{\partial
x^{\nu}} \right) - (\alpha_{l,\dot{l}} (\mu^{2}))_{p\dot{p},
n\dot{n}} \right) = ( - (\partial_{x},
\partial_{x}) - \mu^{2}) \delta_{mn} \delta_{\dot{m} \dot{n}},
\nonumber \\
m,n = 1,...,4l + 2,\, \, \dot{m}, \dot{n} = 1,...,2\dot{l} + 1,
\end{eqnarray}
similar to the relation (\ref{2.28}) is valid. The relations
(\ref{2.33}), (\ref{2.38}) imply that the solution of the equation
(\ref{2.32}) has the form
\begin{eqnarray}
\label{2.39} f_{m\dot{m}} (x) = \sum_{n\, =\, 1}^{4l + 2}
\sum_{\dot{n} \, =\, 1}^{2\dot{l} + 1} \left( \sum_{\nu \, =\,
0}^{3} (\gamma_{l,\dot{l}}^{\nu} (A)\gamma_{l,\dot{l}}^{0}
(A))_{m\dot{m}, n\dot{n}} \left( - i\frac{\partial}{\partial
x^{\nu}} \right) - (\alpha_{l,\dot{l}} (\mu^{2})
\gamma_{l,\dot{l}}^{0} (A))_{m\dot{m}, n\dot{n}} \right) \times
\nonumber \\ \left( - i\int d^{4}ye_{\mu^{2}} (x - y)\delta (y^{0})
f_{n\dot{n}} (+ 0,{\bf y})\right), \,  m = 1,...,4l + 2,\, \dot{m} =
1,...,2\dot{l} + 1.
\end{eqnarray}
We suppose that the smooth function $f_{n\dot{n}} (+ 0,{\bf x})$ is
rapidly decreasing at the infinity. For the solution of the equation
(\ref{2.22}) in the domain $x^{0} < 0$ it is sufficient to use the
distribution $- e_{\mu^{2}} (- x)$ in the relation (\ref{2.39}). By
shifting the integration path in the integral (\ref{2.37}) we have
\begin{eqnarray}
\label{2.40}  \int d^{4}ye_{\mu^{2}} (x - y)\delta (y^{0})
f_{n\dot{n}} (+ 0,{\bf y}) = \nonumber \\ (2\pi)^{- 4} \int
d^{4}p\exp \{ x^{0} - i(p,x)\} ((p^{0} + i)^{2} - |{\bf p}|^{2} -
\mu^{2})^{- 1}
\widetilde{f_{m\dot{m}}} (+ 0,\cdot ) ({\bf p}), \nonumber \\
\widetilde{f_{m\dot{m}}} (+ 0,\cdot ) ({\bf p}) = \int d^{3}{\bf x}
\exp \{ - i\sum_{k\, =\, 1}^{3} p^{k}x^{k}\} f_{m\dot{m}} (+ 0,{\bf
x}).
\end{eqnarray}
The integral with respect to $p^{0}$ may be easily calculated. For
$x^{0} > 0$ and $ \widetilde{f_{m\dot{m}}} (+ 0,\cdot )({\bf p}) =
f_{m\dot{m}} \delta ({\bf p} - {\bf q}) $ the functions
(\ref{2.39}), (\ref{2.40}) are not the eigenfunctions of the
differential operator $ - i\partial /\partial x^{0} $ and are the
eigenfunctions of the differential operator $\left( - i\partial
/\partial x^{0} \right)^{2}$ (see the Dirac discussion of the
negative energy electrons).

Let us introduce the interaction coefficients into the equation
(\ref{2.32}). Let $j,k$ be the permutation of the numbers $1,2$. We
construct the equation for $j$ particle. Let us multiply the
equations (\ref{2.32}) for the particles $1$ and $2$. We change the
differential operator
\begin{equation}
\label{2.41} \left( - i\frac{\partial}{\partial
x_{1}^{\nu_{1}}}\right) \left( - i\frac{\partial}{\partial
x_{2}^{\nu_{2}}}\right)
\end{equation}
for the differential operator
\begin{equation}
\label{2.42} \left( - i\frac{\partial}{\partial
x_{1}^{\nu_{1}}}\right) \left( - i\frac{\partial}{\partial
x_{2}^{\nu_{2}}}\right) + A_{\nu_{1} \nu_{2}}^{(jk)} (x_{j} -
x_{k}).
\end{equation}
The differential operator (\ref{2.42}) should transform like the
differential operator (\ref{2.41}). Therefore for any matrix $A \in
SL(2,{\bf C})$
\begin{equation}
\label{2.43} A_{\nu_{1} \nu_{2}}^{(jk)} \left( \sum_{\mu \, =\,
0}^{3} \Lambda_{\mu}^{\lambda} (A^{- 1})x^{\mu} \right) =
\sum_{\mu_{1}, \mu_{2} \, =\, 0}^{3} \Lambda_{\nu_{1}}^{\mu_{1}}
(A)\Lambda_{\nu_{2}}^{\mu_{2}} (A)A_{\mu_{1} \mu_{2}}^{(jk)} (x).
\end{equation}

The interaction coefficients in the relativistic Coulomb law
(\cite{1}, relations (2.15), (2.16), (2.23)) are defined by the
trajectory of another particle. A particle has no trajectory in the
quantum mechanics. We integrate the obtained equation with respect
to the variable $x_{k}$
\begin{eqnarray}
\label{2.44} \sum_{n_{s} \, =\, 1,...,4l_{s} + 2,\, \, s\, =\, 1,2}
\sum_{\dot{n}_{s} \, =\, 1,...,2\dot{l}_{s} + 1,\, \, s\, =\, 1,2}
\int d^{4}x_{k} \nonumber \\ \{ \prod_{s\, =\, 1}^{2} \left(
\sum_{\nu \, = \, 0}^{3} (\gamma_{l_{s},\dot{l}_{s}}^{\nu}
(A))_{m_{s}\dot{m}_{s}, n_{s}\dot{n}_{s}} \left( -
i\frac{\partial}{\partial x_{s}^{\nu}}\right) +
(\beta_{l_{s},\dot{l}_{s}} (\mu_{s}^{2}))_{m_{s}\dot{m}_{s},
n_{s}\dot{n}_{s}} \right) (f_{s})_{n_{s}\dot{n_{s}}} (x_{s}) +
\nonumber \\ \prod_{s\, =\, 1}^{2} \delta
(x_{s}^{0})(f_{s}^{0})_{n_{s}\dot{n}_{s}} (+ 0,{\bf x}_{s}) +
\nonumber \\ \sum_{\nu_{1}, \nu_{2} \, =\, 0}^{3} A_{\nu_{1}
\nu_{2}}^{(jk)} (x_{j} - x_{k}) \prod_{s\, =\, 1}^{2}
(\gamma_{l_{s},\dot{l}_{s}}^{\nu_{s}} (A))_{m_{s}\dot{m}_{s},
n_{s}\dot{n}_{s}} (f_{s})_{n_{s}\dot{n_{s}}} (x_{s})\} = 0.
\end{eqnarray}
The first and the second terms in the left - hand side of the
equation (\ref{2.44}) correspond to the multiplied equations
(\ref{2.32}) for the particles $1$ and $2$. The third term in the
left - hand side of the equation (\ref{2.44}) corresponds to the
interaction. The equation (\ref{2.44}) transforms like two equations
(\ref{2.32}).

We will construct the interaction coefficients $A_{\nu_{1}
\nu_{2}}^{(jk)} (x)$ using the Clebsch - Gordan coefficients. Due to
(\cite{5}, Chapter III, Section 8.3) we have for a matrix $A \in
SU(2)$
\begin{eqnarray}
\label{2.45} t_{m_{1}n_{1}}^{l_{1}}(A)t_{m_{2}n_{2}}^{l_{2}}(A) =
\sum_{l_{3} \in 1/2{\bf Z}_{+}}
\sum_{m_{3},n_{3} = - l_{3}}^{l_{3}} \nonumber \\
C(l_{1},l_{2},l_{3};m_{1},m_{2},m_{3})C(l_{1},l_{2},l_{3};n_{1},n_{2},n_{3})t_{m_{3}n_{3}}^{l_{3}}(A)
\end{eqnarray}
The Clebsch - Gordan coefficient
$C(l_{1},l_{2},l_{3};m_{1},m_{2},m_{3})$ is not zero only if $m_{3}
= m_{1} + m_{2}$ and the half - integers $l_{1},l_{2},l_{3} \in
1/2{\bf Z}_{+}$ satisfy the triangle condition: the half - integer
$l_{3}$ is one of the half - integers $|l_{1} - l_{2}|,|l_{1} -
l_{2}| + 1,...,l_{1} + l_{2} - 1,l_{1} + l_{2}$. Let the half -
integers $l_{1},l_{2},l_{3} \in 1/2{\bf Z}_{+}$ satisfy the triangle
condition. Let the half - integers $m_{i} = - l_{i}, - l_{i} +
1,...,l_{i} - 1,l_{i}$, $i = 1,2,3$, $m_{3} = m_{1} + m_{2}$. Then
due to (\cite{5}, Chapter III, Section 8.3)
\begin{eqnarray}
\label{2.46} C(l_{1},l_{2},l_{3};m_{1},m_{2},m_{3}) = (- 1)^{l_{1} -
l_{3} + m_{2}} (2l_{3} +
1)^{1/2} \times \nonumber \\
 \left( \frac{(l_{1} + l_{2} - l_{3})!(l_{1} + l_{3} - l_{2})!(l_{2} + l_{3} -
l_{1})!(l_{3} - m_{3})!(l_{3} + m_{3})!}{(l_{1} + l_{2} + l_{3} + 1)!(l_{1} -
m_{1})!(l_{1} + m_{1})!(l_{2} - m_{2})!(l_{2} + m_{2})!}\right)^{1/2}
\times \nonumber \\
\sum_{j = 0}^{l_{2} + l_{3} - l_{1}} \frac{(- 1)^{j}(l_{1} + m_{1} +
j)!(l_{2} + l_{3} - m_{1} - j)!}{j! \Gamma (l_{3} - m_{3} - j + 1)
\Gamma (l_{1} - l_{2} + m_{3} + j + 1)(l_{2} + l_{3} - l_{1} - j)!}.
\end{eqnarray}
Let $dA$ be the normalized Haar measure on the group $SU(2)$. Due to
(\cite{5}, Chapter III, Section 8.3)
\begin{eqnarray}
\label{2.47}
C(l_{1},l_{2},l_{3};m_{1},m_{2},m_{3})C(l_{1},l_{2},l_{3};n_{1},n_{2},n_{3})
= \nonumber \\ (2l_{3} + 1)\int_{SU(2)} dA
t_{m_{1}n_{1}}^{l_{1}}(A)t_{m_{2}n_{2}}^{l_{2}}(A)\overline{t_{m_{3}n_{3}}^{l_{3}}(A)}.
\end{eqnarray}
The coefficients of the polynomial (\ref{2.5}) are real. By using
the relations (\ref{2.7}) and $A^{\ast} = A^{- 1}$ we can rewrite
the equality (\ref{2.47}) as
\begin{eqnarray}
\label{2.48}
C(l_{1},l_{2},l_{3};m_{1},m_{2},m_{3})C(l_{1},l_{2},l_{3};n_{1},n_{2},n_{3})
= \nonumber \\ (2l_{3} + 1)\int_{SU(2)} dA
t_{m_{1}n_{1}}^{l_{1}}(A)t_{m_{2}n_{2}}^{l_{2}}(A)t_{n_{3}m_{3}}^{l_{3}}(A^{-
1}).
\end{eqnarray}
If the half - integers $l_{1},l_{2},l_{3} \in 1/2{\bf Z}_{+}$
satisfy the triangle condition, then due to (\cite{5}, Chapter III,
Section 8.3) we have
\begin{equation}
\label{2.49} C(l_{1},l_{2},l_{3};l_{1},- l_{2},l_{1} - l_{2}) =
\left( \frac{(2l_{3} + 1)(2l_{1})!(2l_{2})!}{(l_{1} + l_{2} -
l_{3})!(l_{1} + l_{2} + l_{3} + 1)!}\right)^{1/2}.
\end{equation}
Let us choose the half - integers $n_{1} = l_{1}, n_{2} = - l_{2}$
in the equality (\ref{2.48}). Then the relations (\ref{2.6}),
(\ref{2.48}), (\ref{2.49}) and the invariance of the Haar measure
$dA$ imply
\begin{eqnarray}
\label{2.50} \sum_{n_{1} = - l_{1}}^{l_{1}} \sum_{n_{2} = -
l_{2}}^{l_{2}}
t_{m_{1}n_{1}}^{l_{1}}(A)t_{m_{2}n_{2}}^{l_{2}}(A)C(l_{1},l_{2},l_{3};n_{1},n_{2},m_{3})
= \nonumber \\ \sum_{n_{3} = - l_{3}}^{l_{3}}
C(l_{1},l_{2},l_{3};m_{1},m_{2},n_{3})t_{n_{3}m_{3}}^{l_{3}}(A).
\end{eqnarray}
The substitution of the matrix $A^{T}$ into the equality
(\ref{2.50}) and the equality (\ref{2.7}) yield
\begin{eqnarray}
\label{2.51} \sum_{n_{1} = - l_{1}}^{l_{1}} \sum_{n_{2} = -
l_{2}}^{l_{2}}
t_{n_{1}m_{1}}^{l_{1}}(A)t_{n_{2}m_{2}}^{l_{2}}(A)C(l_{1},l_{2},l_{3};n_{1},n_{2},m_{3})
= \nonumber \\ \sum_{n_{3} = - l_{3}}^{l_{3}}
C(l_{1},l_{2},l_{3};m_{1},m_{2},n_{3})t_{m_{3}n_{3}}^{l_{3}}(A).
\end{eqnarray}
The relations (\ref{2.50}), (\ref{2.51}) have an analytic
continuation to the group $SL(2,{\bf C})$.

By making use of the matrices (\ref{1.3}) as the coefficients we
define $2\times 2$ - matrix
\begin{equation}
\label{2.52} \tilde{\partial}_{x} = \sum_{\mu = 0}^{3} \eta^{\mu
\mu} \sigma^{\mu} \frac{\partial}{\partial x^{\mu}}.
\end{equation}
We insert the matrix (\ref{2.52}) into the polynomial (\ref{2.5})
and obtain the differential operator
$t_{mn}^{l}(\tilde{\partial}_{x} )$. Due to (\cite{6}, relation (1 -
18)) for any matrix $A \in SL(2,{\bf C})$
\begin{equation}
\label{2.53} \sigma^{2} A^{T}\sigma^{2} = A^{- 1}.
\end{equation}

Similar to the relativistic Coulomb law (\cite{1}, relations (2.15),
(2.16), (2.23)) we construct the interaction coefficients
$A_{\nu_{1} \nu_{2}}^{(jk)} (x)$ from the derivatives of the
distributions (\ref{2.34})
\begin{eqnarray}
\label{2.54} A_{\nu_{1} \nu_{2}}^{(jk)} (x) = K_{jk;0}Q_{\nu_{1}
\nu_{2}}^{0} (\tilde{\partial_{x}}) e_{0}(x) + K_{jk;1}Q_{\nu_{1}
\nu_{2}}^{1} (\tilde{\partial_{x}}) e_{0,0}(x), \nonumber \\
Q_{\nu_{1} \nu_{2}}^{l} (\tilde{\partial_{x}}) =
\sum_{m_{1},m_{2},\dot{m}_{1}, \dot{m}_{2} \, =\, - \frac{1}{2},
\frac{1}{2}} \sum_{m_{12},\dot{m}_{12} \, =\, - l, - l + 1,...,l -
1,l} C(\frac{1}{2}, \frac{1}{2}, l;m_{1},m_{2},m_{12})\times
\nonumber \\ C(\frac{1}{2}, \frac{1}{2}, l;\dot{m}_{1}, \dot{m}_{2},
\dot{m}_{12}) \left( \prod_{k\, =\, 1}^{2}
t_{m_{k}\dot{m}_{k}}^{\frac{1}{2}} (\sigma^{2} \sigma^{\nu_{k}}
\sigma^{2}) \right) t_{m_{12}\dot{m}_{12}}^{l}
(\tilde{\partial_{x}}).
\end{eqnarray}
Here $K_{jk;l}$, $l = 0,1$, are the interaction constants. In view
of the relations (\ref{2.18}), (\ref{2.50}) - (\ref{2.53}) the
interaction coefficients (\ref{2.54}) satisfy the covariance
relation (\ref{2.43}). Due to the triangle condition the Clebsch -
Gordan coefficient $C(\frac{1}{2}, \frac{1}{2},
l;m_{1},m_{2},m_{12})$ is not zero for the integers $l = 0,1$ only.
The degree of the homogeneous polynomial (\ref{2.5}) is $2l$.
Therefore the relation (\ref{2.36}) implies that the support of the
distribution $ t_{m_{12}\dot{m}_{12}}^{1} (\tilde{\partial_{x}})
e_{0,0}(x) $ lies in the boundary of the upper light cone and the
support of the distribution $ t_{m_{12}\dot{m}_{12}}^{1}
(\tilde{\partial_{x}}) e_{0,0,0}(x)$ lies in all closed upper light
cone. Let us prove
\begin{equation}
\label{2.55} Q_{\nu_{1} \nu_{2}}^{0} (\tilde{\partial_{x}}) =
\eta_{\nu_{1} \nu_{2}}.
\end{equation}
The definition (\ref{2.5}) implies $t_{00}^{0}(A) = 1$. Therefore
the left - hand side of the equality (\ref{2.55}) does not depend on
the differential operator (\ref{2.52}). The definition (\ref{1.3})
and the relations (\ref{2.17}), (\ref{2.46}) imply
\begin{equation}
\label{2.56} C(\frac{1}{2}, \frac{1}{2}, 0;m_{1},m_{2},0) = - i2^{-
\frac{1}{2}}t_{m_{1}m_{2}}^{\frac{1}{2}} (\sigma^{2}).
\end{equation}
In view of the definition (\ref{1.3}) $(\sigma^{2})^{T} = -
\sigma^{2}$. The substitution of the relation (\ref{2.56}) into the
definition (\ref{2.54}) yields
\begin{equation}
\label{2.57} Q_{\nu_{1} \nu_{2}}^{0} (\tilde{\partial_{x}}) =
\frac{1}{2} \sum_{m\, =\, - \frac{1}{2}, \frac{1}{2}}
t_{mm}^{\frac{1}{2}} (\sigma^{\nu_{2}} \sigma^{2}
(\sigma^{\nu_{1}})^{T} \sigma^{2}).
\end{equation}
Due to (\cite{6}, Section 1 - 3)
\begin{equation}
\label{2.58} \sigma^{2} (\sigma^{\nu})^{T} \sigma^{2} = \eta_{\nu
\nu} \sigma^{\nu}, \, \, \nu = 0,...,3.
\end{equation}
The relations (\ref{1.3}), (\ref{1.5}), (\ref{2.17}), (\ref{2.57}),
(\ref{2.58}) imply the relation (\ref{2.55}).

For a vector $p \in {\bf R}^{4}$, $(p,p) > 0$, we define a matrix
$(p,p)^{- 1/2} \tilde{p} \in SL(2,{\bf C})$. In view of the
relations (\ref{1.5}) it satisfies the equation
$$
((p,p)^{- 1/2}\tilde{p})^{- 1} = (p,p)^{- 1/2} \sum_{\mu \, =\,
0}^{3} \eta_{\mu \mu} p^{\mu} \sigma^{\mu}.
$$
This relation and the relations (\ref{2.50}), (\ref{2.53}),
(\ref{2.54}) imply
\begin{eqnarray}
\label{2.59} Q_{\nu_{1} \nu_{2}}^{1} ((p,p)^{- 1/2}\tilde{p}) =
\sum_{m_{1},m_{2},\dot{m}_{1}, \dot{m}_{2} \, =\, - \frac{1}{2},
\frac{1}{2}} \sum_{m_{12} \, =\, - 1,0,1} C(\frac{1}{2},
\frac{1}{2}, 1;m_{1},m_{2},m_{12})\times \nonumber \\ C(\frac{1}{2},
\frac{1}{2}, 1;\dot{m}_{1}, \dot{m}_{2}, m_{12}) \prod_{k\, =\,
1}^{2} t_{m_{k}\dot{m}_{k}}^{\frac{1}{2}} \left( \sigma^{2} (p,p)^{-
1/2}\left( \sum_{\mu \, =\, 0}^{3} \eta_{\mu \mu} p^{\mu}
\sigma^{\mu} \right) \sigma^{\nu_{k}} \sigma^{2} \right).
\end{eqnarray}
In view of the relations (\ref{2.10}), (\ref{2.56}) the substitution
of the matrix $A = \sigma^{0}$ into the relation (\ref{2.45}) yields
\begin{eqnarray}
\label{2.60} \sum_{m_{12} \, =\, - 1,0,1} C(\frac{1}{2},
\frac{1}{2}, 1;m_{1},m_{2},m_{12})C(\frac{1}{2}, \frac{1}{2},
1;\dot{m}_{1}, \dot{m}_{2}, m_{12}) = \nonumber
\\ \delta_{m_{1}\dot{m}_{1}} \delta_{m_{2}\dot{m}_{2}} + \frac{1}{2}
t_{m_{1}m_{2}}^{\frac{1}{2}} (\sigma^{2}) t_{\dot{m}_{1}
\dot{m}_{2}}^{\frac{1}{2}} (\sigma^{2}).
\end{eqnarray}
The relations (\ref{1.3}), (\ref{1.5}), (\ref{2.17}), (\ref{2.53}),
(\ref{2.58}) - (\ref{2.60}) imply
$$
Q_{\nu_{1} \nu_{2}}^{1} ((p,p)^{- 1/2}\tilde{p}) = 4(p,p)^{-
1}\eta_{\nu_{1} \nu_{1}} \eta_{\nu_{2} \nu_{2}} p^{\nu_{1}}
p^{\nu_{2}} - \eta_{\nu_{1} \nu_{2}}.
$$
This relation and the definitions (\ref{2.5}), (\ref{2.52}),
(\ref{2.54}) imply
\begin{equation}
\label{2.61} Q_{\nu_{1} \nu_{2}}^{1} (\tilde{\partial_{x}}) =
4\frac{\partial}{\partial x^{\nu_{1}}} \frac{\partial}{\partial
x^{\nu_{2}}} - \eta_{\nu_{1} \nu_{2}} (\partial_{x}, \partial_{x}).
\end{equation}
The substitution of the relations (\ref{2.55}), (\ref{2.61}) into
the definitions (\ref{2.37}), (\ref{2.54}) yields
\begin{eqnarray}
\label{2.62} A_{\nu_{1} \nu_{2}}^{(jk)} (x) = \lim_{\epsilon
\rightarrow + 0} (K_{jk;0} + K_{jk;1})(2\pi)^{- 4} \int d^{4}p\exp
\{ - i(p,x)\} \times \nonumber \\ ((p^{0} + i\epsilon)^{2} - |{\bf
p}|^{2})^{- 1} \left( \eta_{\nu_{1} \nu_{2}} -
\frac{4K_{jk;1}}{K_{jk;0} + K_{jk;1}} \frac{p_{\nu_{1}} p_{\nu_{2}}
}{(p^{0} + i\epsilon)^{2} - |{\bf p}|^{2}} \right).
\end{eqnarray}
Due to (\cite{7}, Chapter III, relation (3.58)) the propagation
function of the vector particles in the Yang - Mills theory is
\begin{eqnarray}
\label{2.63} D_{\mu \nu}^{ab} (x) = \lim_{\epsilon \rightarrow + 0}
- \delta^{ab} (2\pi)^{- 4} \int d^{4}p\exp \{ - i(p,x)\} \times
\nonumber \\ ((p,p) + i\epsilon)^{- 1} \left( \eta_{\mu \nu} - (1 -
\alpha )\frac{p_{\mu} p_{\nu}}{(p,p) + i\epsilon }\right).
\end{eqnarray}
The number $\alpha$ is the consequence of the gauge condition. The
choices $\alpha = 1$ and $\alpha = 0$ are called the gauge
conditions of Feynman and Landau. The distribution (\ref{2.63})
differs from the distribution (\ref{2.62}) in the rule of going
around the poles in the integral.

In the quantum electrodynamics (\cite{8}, Lecture 24) the equation
\begin{eqnarray}
\label{2.64} \sum_{n_{1},n_{2}\, =\, 1}^{4} \{ \left( \prod_{s\, =\,
1}^{2} \left( \sum_{\nu \, =\, 0}^{3} (\gamma_{\frac{1}{2}, 0}^{\nu}
(\sigma^{0}))_{m_{s}0,n_{s}0} \left( i\frac{\partial}{\partial
x_{s}^{\nu}}\right) - \mu_{s} \delta_{m_{s}n_{s}} \right) \right)
\psi_{n_{1}n_{2},p_{1}p_{2}} (x_{1},x_{2}) +  \sum_{\nu_{1}, \nu_{2}
\, =\, 0}^{3} \nonumber \\ \eta_{\nu_{1} \nu_{2}}
K_{0}D_{0}^{c}(x_{1} - x_{2})\left( \prod_{s\, =\, 1}^{2}
(\gamma_{\frac{1}{2}, 0}^{\nu_{s}}
(\sigma^{0}))_{m_{s}0,n_{s}0}\right) \psi_{n_{1}n_{2},p_{1}p_{2}}
(x_{1},x_{2})\} = i\prod_{s\, =\, 1}^{2} \delta (x_{s})
\delta_{m_{s}p_{s}}, \nonumber \\ D_{0}^{c}(x) = \lim_{\epsilon
\rightarrow + 0} - 2(2\pi)^{- 3} \int d^{4}p\exp \{ - i(p,x)\}
((p,p) + i\epsilon)^{- 1}
\end{eqnarray}
is studied. In view of the relation (\ref{2.37}) the distribution $-
(4\pi)^{- 1}D_{0}^{c}(x)$ differs from the distribution $e_{0}(x)$
in the rule of going around the poles in the integral. $
D_{0}^{c}(x_{2} - x_{1}) = D_{0}^{c} (x_{1} - x_{2})$, $e_{0}(x_{2}
- x_{1}) \neq e_{0}(x_{1} - x_{2})$. Therefore the equation
(\ref{2.44}), (\ref{2.54}), (\ref{2.55}), $K_{jk;1} = 0$ without
integration with respect to the variable $x_{k}$ has the logical
contradiction. The right - hand sides of the equations (\ref{2.44})
and (\ref{2.64}) are similar if at the initial moment both particles
are concentrated at the origin of coordinates.

We have considered up to now that the interaction propagates at the
speed of light. Let the interaction propagate at the speed less or
equal to the speed of light. Hence the interaction coefficients
(\ref{2.54}) may be changed in the following way
\begin{equation}
\label{2.65} A_{\nu_{1} \nu_{2}}^{(jk)} (x) = \sum_{l_{12}\, \in \,
1/2{\bf Z}_{+}} \int d\lambda_{1} \cdots d\lambda_{l_{12} + 1}
K_{jk;l_{12}}(\lambda_{1},...,\lambda_{l_{12} + 1}) Q_{\nu_{1}
\nu_{2}}^{l_{12}} (\tilde{\partial}_{x})
e_{\lambda_{1}^{2},...,\lambda_{l_{12} + 1}^{2}} (x).
\end{equation}
The distributions $K_{jk;l_{12}}(\lambda_{1},...,\lambda_{l_{12} +
1})$ have the compact supports. In view of the relations
(\ref{2.18}), (\ref{2.50}) - (\ref{2.53}) the interaction
coefficients (\ref{2.65}) satisfy the covariance relation
(\ref{2.43}).

Let us consider the equation (\ref{2.44}), (\ref{2.65}), $j = 1$ ,
$k = 2$ for the odd integers $2l_{1} + 2\dot{l}_{1}$, $2l_{2} +
2\dot{l}_{2}$. Let the functions $(f_{2})_{m_{2}\dot{m}_{2}}
(x_{2})$ be the solutions of the equation (\ref{2.32}). The heavy
second particle moves freely. The functions
$(f_{2})_{m_{2}\dot{m}_{2}} (x_{2})$  are given  by the relations
(\ref{2.39}), (\ref{2.40}). Let the initial function
$(f_{2})_{m_{2}\dot{m}_{2}} (+ 0,{\bf x}_{2})$ be such that the
integral of the function (\ref{2.66}) is not equal to zero for some
numbers $m_{2},\dot{m}_{2}$. In view of the relation (\ref{2.39})
the equation (\ref{2.44}), $j = 1$, $k = 2$ implies
\begin{eqnarray}
\label{2.67} \sum_{n_{1}\, =\, 1}^{4l_{1} + 2} \sum_{\dot{n}_{1} \,
=\, 1}^{2\dot{l}_{1} + 1} \left( \sum_{\nu \, =\, 0}^{3}
(\gamma_{l_{1},\dot{l}_{1}}^{\nu} (A))_{m_{1}\dot{m}_{1},
n_{1}\dot{n}_{1}} \left( - i\frac{\partial}{\partial x_{1}^{\nu}}
\right) + (\beta_{l_{1},\dot{l}_{1}}
(\mu_{1}^{2}))_{m_{1}\dot{m}_{1}, n_{1}\dot{n}_{1}} \right)
(f_{1})_{n_{1}\dot{n}_{1}} (x_{1}) + \nonumber \\ i\left( \int
d^{3}{\bf y}_{2} (f_{2}^{0})_{m_{2}\dot{m}_{2}} (+ 0,{\bf y}_{2})
\right)^{- 1} \sum_{\nu_{1}, \nu_{2} \, =\, 0}^{3} \int d^{4}x_{2}
A_{\nu_{1} \nu_{2}}^{(12)} (x_{1} - x_{2}) \times \nonumber
\\  \left( \prod_{k\, =\, 1,2} \left( \sum_{n_{k}\, =\, 1}^{4l_{k} +
2} \sum_{\dot{n}_{k} \, =\, 1}^{2\dot{l}_{k} + 1}
(\gamma_{l_{k},\dot{l}_{k}}^{\nu_{k}} (A))_{m_{k}\dot{m}_{k},
n_{k}\dot{n}_{k}} (f_{k})_{n_{k}\dot{n}_{k}} (x_{k})\right) \right)
= - i\delta (x_{1}^{0})
(f_{1}^{0})_{m_{1}\dot{m}_{1}} (+ 0,{\bf x}_{1}), \\
\sum_{n_{2}\, =\, 1}^{4l_{2} + 2} \sum_{\dot{n}_{2} \, =\,
1}^{2\dot{l}_{2} + 1} (\gamma_{l_{2},\dot{l}_{2}}^{\nu_{2}}
(A))_{m_{2}\dot{m}_{2}, n_{2}\dot{n}_{2}} (f_{2})_{n_{2}\dot{n}_{2}}
(x_{2}) =  \sum_{n_{2}\, =\, 1}^{4l_{2} + 2} \sum_{\dot{n}_{2} \,
=\, 1}^{2\dot{l}_{2} + 1} \int d^{4}x_{3} \nonumber
\\ (P_{2}^{\nu_{2}})_{m_{2}\dot{m}_{2}, n_{2}\dot{n}_{2}} \left(
i\frac{\partial}{\partial x_{2}^{\lambda}} \right) e_{\mu_{2}^{2}}
(x_{2} - x_{3})\delta (x_{3}^{0})(f_{2}^{0})_{n_{2}\dot{n}_{2}} (+
0,{\bf x}_{3}),  \nonumber
\end{eqnarray}
\begin{eqnarray}
\label{2.68} (P_{k}^{\nu_{k}})_{m_{k}\dot{m}_{k}, n_{k}\dot{n}_{k}}
\left( i\frac{\partial}{\partial x_{k}^{\lambda}} \right) =
\sum_{\nu \, =\, 0}^{3} i(\gamma_{l_{k},\dot{l}_{k}}^{\nu_{k}}
(A)\gamma_{l_{k},\dot{l}_{k}}^{\nu} (A))_{m_{k}\dot{m}_{k},
n_{k}\dot{n}_{k}} \left( i\frac{\partial}{\partial x_{k}^{\nu}}
\right) + \nonumber \\ i(\gamma_{l_{k},\dot{l}_{k}}^{\nu_{k}}
(A)\alpha_{l_{k}\dot{l}_{k}} (\mu_{k}^{2}))_{m_{k}\dot{m}_{k},
n_{k}\dot{n}_{k}}, \, \, \nu_{k} = 0,...,3,\, \, k = 1,2.
\end{eqnarray}
The equation (\ref{2.67}) has the simple physical meaning: the
substitution of the first relation (\ref{2.34}) into the equation
(\ref{2.54}), (\ref{2.55}), (\ref{2.67}), $K_{12;1} = 0$ yields the
equation (\ref{2.32}) with the retarded potential of the second
particle. We have inserted into the equation (\ref{2.67}) the
functions $(f_{2})_{m_{2}\dot{m}_{2}} (x_{2})$ given by the
relations (\ref{2.39}), (\ref{2.40}). We suppose that the functions
$\widetilde{(f_{2})}_{m_{2}\dot{m_{2}}} (+ 0,\cdot) ({\bf p}_{2})$
are rapidly decreasing at the infinity. It interesting to study also
the solutions of the equation (\ref{2.67}) for the functions
$(f_{2})_{m_{2}\dot{m}_{2}}( + 0,{\bf x}_{2}) =
(c_{2})_{m_{2}\dot{m}_{2}}\delta ({\bf x}_{2})$. For the functions
$(f_{2})_{m_{2}\dot{m}_{2}} (x_{2}) = (c_{2})_{m_{2}\dot{m}_{2}}
\delta ({\bf x}_{2})$ the equation (\ref{2.54}), (\ref{2.55}),
(\ref{2.67}), $K_{12;1} = 0$ is the equation (\ref{2.32}) with the
Coulomb potential. This equation has the discrete energy levels.

Similar to the solutions (\ref{2.39}) of the equation (\ref{2.32})
we look for the solutions of the equation (\ref{2.67}) in the form
\begin{eqnarray}
\label{2.69} (f_{1})_{m_{1}\dot{m}_{1}} (x_{1}) = \sum_{n_{1}\, =\,
1}^{4l_{1} + 2} \sum_{\dot{n}_{1} \, =\, 1}^{2\dot{l}_{1} + 1}
\nonumber \\ \left( \sum_{\nu \, =\, 0}^{3}
(\gamma_{l_{1},\dot{l}_{1}}^{\nu} (A))_{m_{1}\dot{m}_{1},
n_{1}\dot{n}_{1}} \left( - i\frac{\partial}{\partial x_{1}^{\nu}}
\right) - (\alpha_{l_{1},\dot{l}_{1}}
(\mu_{1}^{2}))_{m_{1}\dot{m}_{1}, n_{1}\dot{n}_{1}} \right)
(g_{1})_{n_{1}\dot{n}_{1}} (x).
\end{eqnarray}
In view of the relations (\ref{2.33}), (\ref{2.38}) the substitution
of the function (\ref{2.69}) into the equation (\ref{2.67}) and the
convolution of the obtained equation with the distribution
$e_{\mu_{1}^{2}} (x)$ yield
\begin{eqnarray}
\label{2.70} (g_{1})_{m_{1}\dot{m}_{1}} (x_{1}) - \sum_{n_{s} \, =\,
1,...,4l_{s} + 2,\, \, s\, =\, 1,2} \sum_{\dot{n}_{s} \, =\,
1,...,2\dot{l}_{s} + 1,\, \, s\, =\, 1,2} \sum_{\nu_{1}, \nu_{2} \,
=\, 0}^{3} \int d^{4}x_{2}d^{4}x_{3}d^{4}x_{4} \nonumber
\\ ie_{\mu_{1}^{2}} (x_{1} - x_{2})A_{\nu_{1}
\nu_{2}}^{(12)} (x_{2} - x_{3})(P_{1}^{\nu_{1}})_{m_{1}\dot{m}_{1},
n_{1}\dot{n}_{1}} \left( i\frac{\partial}{\partial x_{2}^{\lambda}}
\right) (g_{1})_{n_{1}\dot{n}_{1}} (x_{2}) \times \nonumber \\
\left( \int d^{3}{\bf y}_{2} (f_{2}^{0})_{m_{2}\dot{m}_{2}} (+
0,{\bf y}_{2}) \right)^{- 1} (P_{2}^{\nu_{2}})_{m_{2}\dot{m}_{2},
n_{2}\dot{n}_{2}} \left( i\frac{\partial}{\partial x_{3}^{\lambda}}
\right) e_{\mu_{2}^{2}} (x_{3} - x_{4})\times \nonumber \\ \delta
(x_{4}^{0})(f_{2}^{0})_{n_{2}\dot{n}_{2}} (+ 0,{\bf x}_{4}) = -
i\int d^{4}x_{2} e_{\mu_{1}^{2}} (x_{1} - x_{2}) \delta (x_{2}^{0})
(f_{1}^{0})_{m_{1}\dot{m}_{1}} (+ 0,{\bf x}_{2}).
\end{eqnarray}
The relations (\ref{2.37}), (\ref{2.65}) allow to rewrite the
equation (\ref{2.70}) in the form
\begin{eqnarray}
\label{2.71} (g_{1})_{m_{1}\dot{m}_{1}} (x_{1}) - \lim_{\epsilon
\rightarrow + 0} \sum_{n_{s} \, =\, 1,...,4l_{s} + 2,\, \, s\, =\,
1,2} \sum_{\dot{n}_{s} \, =\, 1,...,2\dot{l}_{s} + 1,\, \, s\, =\,
1,2} \sum_{\nu_{1}, \nu_{2} \, =\, 0}^{3} \sum_{l_{12} \, \in \, 1/2
{\bf Z}_{+}} \int d^{4}p_{1}d^{4}p_{2} \nonumber
\\ \int d\lambda_{1} \cdots d\lambda_{l_{12} + 1}
i(2\pi)^{- 8} K_{12;l_{12}}(\lambda_{1},...,\lambda_{l_{12} + 1})
\exp \{ - i(x_{1},p_{1})\} \times \nonumber \\ ((p_{1}^{0} +
i\epsilon)^{2} - |{\bf p}_{1}|^{2} - \mu_{1}^{2})^{- 1} ((p_{1}^{0}
- p_{2}^{0} + i\epsilon)^{2} - |{\bf p}_{1} - {\bf p}_{2}|^{2} -
\mu_{1}^{2})^{- 1} \times \nonumber \\ \left( \prod_{j\, =\,
1}^{l_{12} + 1} ((p_{1}^{0} - p_{2}^{0} + i\epsilon)^{2} - |{\bf
p}_{1} - {\bf p}_{2}|^{2} - \lambda_{j}^{2})^{- 1} \right)
Q_{\nu_{1} \nu_{2}}^{l_{12}} (- i(\tilde{p}_{1} - \tilde{p}_{2}))
\times \nonumber \\ (P_{1}^{\nu_{1}})_{m_{1}\dot{m}_{1},
n_{1}\dot{n}_{1}} (\eta_{\nu \nu} p_{2}^{\nu})
(\tilde{g}_{1})_{n_{1}\dot{n}_{1}} (p_{2}) \left( \int d^{3}{\bf
y}_{2} (f_{2}^{0})_{m_{2}\dot{m}_{2}} (+ 0,{\bf y}_{2}) \right)^{-
1} \times \nonumber \\ (P_{2}^{\nu_{2}})_{m_{2}\dot{m}_{2},
n_{2}\dot{n}_{2}} (\eta_{\nu \nu} (p_{1}^{\nu} - p_{2}^{\nu}))
\widetilde{(f_{2}^{0})}_{n_{2}\dot{n}_{2}} (+ 0,\cdot) ({\bf p}_{1}
- {\bf p}_{2}) = - i(2\pi)^{- 4}\lim_{\epsilon \, \rightarrow \, +
0} \nonumber
\\ \int d^{4}p_{1}\exp \{ - i(x_{1},p_{1})\} ((p_{1}^{0} +
i\epsilon)^{2} - |{\bf p}_{1}|^{2} - \mu_{1}^{2})^{- 1}
\widetilde{(f_{1}^{0})}_{m_{1}\dot{m}_{1}} (+
0,\cdot) ({\bf p}_{1}), \nonumber \\
(\tilde{g}_{1})_{n_{1}\dot{n}_{1}} (p_{1}) = \int d^{4}x_{1}\exp \{
i(x_{1},p_{1})\} (g_{1})_{n_{1}\dot{n}_{1}} (x_{1}).
\end{eqnarray}
Let us consider the series
\begin{eqnarray}
\label{2.72} (g_{1})_{m_{1}^{(1)} \dot{m}_{1}^{(1)}} (x_{1}) = -
i(2\pi)^{- 4} \lim_{\epsilon \rightarrow + 0} \int d^{4}p_{1} \exp
\{ - i(x_{1},p_{1})\} \times \nonumber \\ ((p_{1}^{0} +
i\epsilon)^{2} - |{\bf p}_{1}|^{2} - \mu_{1}^{2})^{- 1}
\widetilde{(f_{1}^{0})}_{m_{1}^{(1)}\dot{m}_{1}^{(1)}} (+ 0,\cdot)
({\bf p}_{1}) \nonumber \\ - i(2\pi)^{- 4} \lim_{\epsilon
\rightarrow + 0} \sum_{k\, =\, 1}^{\infty} \sum_{m_{1}^{(s)} \, =\,
1,...,4l_{s} + 2,\, \, s\, =\, 2,...,k + 1} \sum_{\dot{m}_{1}^{(s)}
\, =\, 1,...,2\dot{l}_{s} + 1,\, \, s\, =\, 2,...,k + 1} \nonumber
\\ \int d^{4}p_{1}\cdots d^{4}p_{k + 1} \exp \{ - i(x_{1},p_{1})\}
\times \nonumber \\ ((p_{k + 1}^{0} + i\epsilon)^{2} - |{\bf p}_{k +
1}|^{2} - \mu_{1}^{2})^{- 1} \widetilde{(f_{1}^{0})}_{m_{1}^{(k +
1)}\dot{m}_{1}^{(k + 1)}} (+ 0,\cdot) ({\bf p}_{k + 1}) \times
\nonumber \\ \prod_{s\, =\, 1}^{k} \sum_{n_{2}\, =\, 1,...,4l_{2} +
2} \sum_{\dot{n}_{2} \, =\, 1,...,2\dot{l}_{2} + 1} \sum_{\nu_{1},
\nu_{2} \, =\, 0}^{3} \sum_{l_{12} \, \in \, 1/2 {\bf Z}_{+}}
\nonumber \\ \int d\lambda_{1} \cdots d\lambda_{l_{12} + 1}
i(2\pi)^{- 8} K_{12;l_{12}}(\lambda_{1},...,\lambda_{l_{12} + 1})
\times \nonumber
\\ ((p_{s}^{0} + i\epsilon)^{2} - |{\bf p}_{s}|^{2} -
\mu_{1}^{2})^{- 1} ((p_{s}^{0} - p_{s + 1}^{0} + i\epsilon)^{2} -
|{\bf p}_{s} - {\bf p}_{s + 1}|^{2} - \mu_{1}^{2})^{- 1} \times
\nonumber \\ \left( \prod_{j\, =\, 1}^{l_{12} + 1} ((p_{s}^{0} -
p_{s + 1}^{0} + i\epsilon)^{2} - |{\bf p}_{s} - {\bf p}_{s + 1}|^{2}
- \lambda_{j}^{2})^{- 1} \right) Q_{\nu_{1} \nu_{2}}^{l_{12}} (-
i(\tilde{p}_{s} - \tilde{p}_{s +
1})) \times \nonumber \\
(P_{1}^{\nu_{1}})_{m_{1}^{(s)}\dot{m}_{1}^{(s)}, m_{1}^{(s +
1)}\dot{m}_{1}^{(s + 1)}} (\eta_{\nu \nu} p_{s + 1}^{\nu}) \left(
\int d^{3}{\bf y}_{2} (f_{2}^{0})_{m_{2}\dot{m}_{2}}
(+ 0,{\bf y}_{2}) \right)^{- 1} \times \nonumber \\
(P_{2}^{\nu_{2}})_{m_{2}\dot{m}_{2}, n_{2}\dot{n}_{2}} (\eta_{\nu
\nu} (p_{s}^{\nu} - p_{s + 1}^{\nu}))
\widetilde{(f_{2}^{0})}_{n_{2}\dot{n}_{2}} (+ 0,\cdot) ({\bf p}_{s}
- {\bf p}_{s + 1}).
\end{eqnarray}
If all integrals in the right - hand side of the equality
(\ref{2.72}) exist and the series (\ref{2.72}) is convergent, this
series is the solution of the equation (\ref{2.71}). The function of
the variables $p_{1}^{0}$,...,$p_{k + 1}^{0}$ in $k$ term of the
series (\ref{2.72}) is holomorphic in the domain $ \hbox{Im}
p_{1}^{0} > \hbox{Im} p_{2}^{0} > \cdots > \hbox{Im} p_{k + 1}^{0} >
0$. We suppose that the integrals with respect to the variables
$p_{1}^{0}$,...,$p_{k + 1}^{0}$ are absolutely convergent and it is
possible to make the shifts $p_{s}^{0} \rightarrow p_{s}^{0} + i(k +
2 - s)$, $s = 1,...,k + 1$,
\begin{eqnarray}
\label{2.73} (g_{1})_{m_{1}^{(1)} \dot{m}_{1}^{(1)}} (x_{1}) = -
i(2\pi)^{- 4} \int d^{4}p_{1} \exp \{ x_{1}^{0} - i(x_{1},p_{1})\}
((p_{1}^{0} + i)^{2} - |{\bf p}_{1}|^{2}
- \mu_{1}^{2})^{- 1} \times \nonumber \\
\widetilde{(f_{1}^{0})}_{m_{1}^{(1)}\dot{m}_{1}^{(1)}} (+ 0,\cdot)
({\bf p}_{1}) - i(2\pi)^{- 4} \sum_{k\, =\, 1}^{\infty}
\sum_{m_{1}^{(s)} \, =\, 1,...,4l_{s} + 2,\, \, s\, =\, 2,...,k + 1}
\sum_{\dot{m}_{1}^{(s)} \, =\, 1,...,2\dot{l}_{s} + 1,\, \, s\, =\,
2,...,k + 1} \nonumber
\\ \int d^{4}p_{1}\cdots
d^{4}p_{k + 1} \exp \{ x_{1}^{0}(k + 1) - i(x_{1},p_{1})\} ((p_{k +
1}^{0} + i)^{2} - |{\bf p}_{k + 1}|^{2} - \mu_{1}^{2})^{- 1} \times
\nonumber
\\ \widetilde{(f_{1}^{0})}_{m_{1}^{(k + 1)}\dot{m}_{1}^{(k + 1)}} (+
0,\cdot) ({\bf p}_{k + 1}) \prod_{s\, =\, 1}^{k} \sum_{n_{2}\, =\,
1,...,4l_{2} + 2} \sum_{\dot{n}_{2} \, =\, 1,...,2\dot{l}_{2} + 1}
\sum_{\nu_{1}, \nu_{2} \, =\, 0}^{3} \sum_{l_{12} \, \in \, 1/2 {\bf
Z}_{+}} \nonumber \\ ((p_{s}^{0} + i(k + 2 - s))^{2} - |{\bf
p}_{s}|^{2} - \mu_{1}^{2})^{- 1} ((p_{s}^{0} - p_{s + 1}^{0} +
i)^{2} - |{\bf p}_{s} - {\bf p}_{s + 1}|^{2} - \mu_{1}^{2})^{- 1}
\times \nonumber
\\  \int d\lambda_{1} \cdots d\lambda_{l_{12} + 1}
i(2\pi)^{- 8} K_{12;l_{12}}(\lambda_{1},...,\lambda_{l_{12} + 1})
\times \nonumber \\ \left( \prod_{j\, =\, 1}^{l_{12} + 1}
((p_{s}^{0} - p_{s + 1}^{0} + i)^{2} - |{\bf p}_{s} - {\bf p}_{s +
1}|^{2} - \lambda_{j}^{2})^{- 1} \right) Q_{\nu_{1}
\nu_{2}}^{l_{12}} (- i(\tilde{p}_{s} -
\tilde{p}_{s + 1}) + \sigma^{0}) \times \nonumber \\
(P_{1}^{\nu_{1}})_{m_{1}^{(s)}\dot{m}_{1}^{(s)}, m_{1}^{(s +
1)}\dot{m}_{1}^{(s + 1)}} (\eta_{\nu \nu} p_{s + 1}^{\nu} +
i\eta_{\nu 0} (k + 1 - s)) \left( \int d^{3}{\bf y}_{2}
(f_{2}^{0})_{m_{2}\dot{m}_{2}}
(+ 0,{\bf y}_{2}) \right)^{- 1} \times \nonumber \\
(P_{2}^{\nu_{2}})_{m_{2}\dot{m}_{2}, n_{2}\dot{n}_{2}} (\eta_{\nu
\nu} (p_{s}^{\nu} - p_{s + 1}^{\nu}) +i\eta_{\nu 0})
\widetilde{(f_{2}^{0})}_{n_{2}\dot{n}_{2}} (+ 0,\cdot) ({\bf p}_{s}
- {\bf p}_{s + 1}).
\end{eqnarray}
Let us prove that all integrals in the right - hand side of the
equality (\ref{2.73}) are absolutely convergent if the functions
$\widetilde{(f_{k}^{0})}_{m_{k}\dot{m}_{k}} (+ 0,\cdot) ({\bf
p}_{k})$, $k = 1,2$, are rapidly decreasing at the infinity.
Similarly we can prove that it is possible to make the shifts
$p_{s}^{0} \rightarrow p_{s}^{0} + i(k + 2 - s)$, $s = 1,...,k + 1$,
in the $k$ term of the series (\ref{2.72}).

Let us estimate the quadratic polynomial of the variable
$(p^{0})^{2}$
$$
|(p^{0} + im)^{2} - |{\bf p}|^{2} - \mu^{2}|^{2} = ((p^{0})^{2} -
|{\bf p}|^{2} - \mu^{2} + m^{2})^{2} + 4m^{2}(|{\bf p}|^{2} +
\mu^{2}),\, \, m = 1,2,... .
$$
For $|{\bf p}|^{2} + \mu^{2} \geq m^{2}$
\begin{equation}
\label{2.74} |(p^{0} + im)^{2} - |{\bf p}|^{2} - \mu^{2}|^{2} \geq
4m^{2}(|{\bf p}|^{2} + \mu^{2}).
\end{equation}
For $|{\bf p}|^{2} + \mu^{2} \leq m^{2}$
\begin{equation}
\label{2.75} |(p^{0} + im)^{2} - |{\bf p}|^{2} - \mu^{2}|^{2} \geq
(|{\bf p}|^{2} + \mu^{2} + m^{2})^{2}.
\end{equation}
The inequalities (\ref{2.74}), (\ref{2.75}) imply
\begin{equation}
\label{2.76} |(p^{0} + im)^{2} - |{\bf p}|^{2} - \mu^{2}|^{- 1} \leq
m^{- 2},
\end{equation}
\begin{equation}
\label{2.77} |{\bf p}||(p^{0} + im)^{2} - |{\bf p}|^{2} -
\mu^{2}|^{- 1} \leq (2m)^{- 1}.
\end{equation}
By making use of the equality
\begin{equation}
\label{2.78} |(p^{0} + im)^{2} - |{\bf p}|^{2} - \mu^{2}|^{2} =
((p^{0})^{2} - |{\bf p}|^{2} - \mu^{2} - m^{2})^{2} +
4m^{2}(p^{0})^{2}
\end{equation}
we obtain the inequality
\begin{equation}
\label{2.79} |p^{0}||(p^{0} + im)^{2} - |{\bf p}|^{2} - \mu^{2}|^{-
1} \leq (2m)^{- 1}.
\end{equation}
The inequality (\ref{2.76}), the equality (\ref{2.78}) and the
equality
$$
(p^{0})^{2} = ((p^{0})^{2} - |{\bf p}|^{2} - \mu^{2} - m^{2}) +
(|{\bf p}|^{2} + \mu^{2} + m^{2})
$$
imply the inequality
\begin{equation}
\label{2.80} (1 + (p^{0})^{2})|(p^{0} + im)^{2} - |{\bf p}|^{2} -
\mu^{2}|^{- 1} \leq m^{- 2}(|{\bf p}|^{2} + \mu^{2} + 2m^{2} + 1).
\end{equation}
The inequalities (\ref{2.76}), (\ref{2.77}), (\ref{2.79}) imply
\begin{eqnarray}
\label{2.81} |(P_{1}^{\nu_{1}})_{m_{1}^{(s)}\dot{m}_{1}^{(s)},
m_{1}^{(s + 1)}\dot{m}_{1}^{(s + 1)}} (\eta_{\nu \nu} p_{s +
1}^{\nu} + i\eta_{\nu 0} (k + 1 - s))|\times \nonumber \\ |(p_{s +
1}^{0} + i(k + 1 - s))^{2} - |{\bf p}_{s + 1}|^{2} - \mu_{1}^{2}|^{-
1} \leq (k + 1 - s)^{- 1}D_{m_{1}^{(s)}\dot{m}_{1}^{(s)}, m_{1}^{(s
+ 1)}\dot{m}_{1}^{(s + 1)}}^{1,\nu_{1}}, \nonumber \\
|(P_{2}^{\nu_{2}})_{m_{2}\dot{m}_{2}, n_{2}\dot{n}_{2}} (\eta_{\nu
\nu} (p_{s}^{\nu} - p_{s + 1}^{\nu}) +i\eta_{\nu 0})| |(p_{s}^{0} -
p_{s + 1}^{0} + i)^{2} - |{\bf p}_{s} - {\bf p}_{s + 1}|^{2} -
\mu_{1}^{2}|^{- 1} \leq \nonumber \\ D_{m_{2}\dot{m}_{2},
n_{2}\dot{n}_{2}}^{2,\nu_{2}},\, \, \nu_{1}, \nu_{2} = 0,...,3,\, \,
s = 1,...,k
\end{eqnarray}
where the positive matrix elements $
D_{m_{1}^{(s)}\dot{m}_{1}^{(s)}, m_{1}^{(s + 1)}\dot{m}_{1}^{(s +
1)}}^{1,\nu_{1}}$, $D_{m_{2}\dot{m}_{2},
n_{2}\dot{n}_{2}}^{2,\nu_{2}} $ do not depend on the vectors
$p_{s}$, $s = 1,...,k + 1$.

The degree of the homogeneous polynomial (\ref{2.5}) is equal to
$2l$. Hence the inequalities (\ref{2.76}), (\ref{2.80}) imply
\begin{eqnarray}
\label{2.82} |Q_{\nu_{1} \nu_{2}}^{l_{12}} (- i(\tilde{p}_{s} -
\tilde{p}_{s + 1}) + \sigma^{0})| \Biggl| \int d\lambda_{1} \cdots
d\lambda_{l_{12} + 1} K_{12;l_{12}}(\lambda_{1},...,\lambda_{l_{12}
+ 1}) \times \nonumber \\ \left( \prod_{j\, =\, 1}^{l_{12} + 1}
((p_{s}^{0} - p_{s + 1}^{0} + i)^{2} - |{\bf p}_{s} - {\bf p}_{s +
1}|^{2} - \lambda_{j}^{2})^{- 1} \right) \Biggr| \leq D_{\nu_{1}
\nu_{2}}^{3,l_{12}} (1 + (p_{s}^{0} - p_{s + 1}^{0})^{2})^{- 1}
\times \nonumber \\ (|{\bf p}_{s} - {\bf p}_{s + 1}|^{2} + \max
\lambda_{j}^{2} + 3)^{2},\, \, \nu_{1}, \nu_{2} = 0,...,3,\, \, s =
1,...,k
\end{eqnarray}
where the positive matrix elements $D_{\nu_{1} \nu_{2}}^{3,l_{12}}$
do not depend on the vectors $p_{s}$, $s = 1,...,k + 1$.

The inequalities (\ref{2.81}), (\ref{2.82}) imply the estimation of
the series (\ref{2.73})
\begin{eqnarray}
\label{2.83} |(g_{1})_{m_{1}^{(1)} \dot{m}_{1}^{(1)}} (x_{1})| \leq
(2\pi)^{- 4} \int d^{4}p_{1} \exp \{ x_{1}^{0} \} |(p_{1}^{0} +
i)^{2} - |{\bf p}_{1}|^{2} - \mu_{1}^{2}|^{- 1} \times \nonumber
\\ |\widetilde{(f_{1}^{0})}_{m_{1}^{(1)}\dot{m}_{1}^{(1)}} (+ 0,\cdot)
({\bf p}_{1})|  +  (2\pi)^{- 4} \sum_{k\, =\, 1}^{\infty}
\frac{1}{k!} \int d^{4}p_{1}\cdots d^{4}p_{k + 1} \exp \{
x_{1}^{0}(k + 1)\} \times \nonumber \\ \sum_{m_{1}^{(s)} \, =\,
1,...,4l_{s} + 2,\, \, s\, =\, 2,...,k + 1} \sum_{\dot{m}_{1}^{(s)}
\, =\, 1,...,2\dot{l}_{s} + 1,\, \, s\, =\, 2,...,k + 1} |(p_{1}^{0}
+ i(k + 1))^{2} - |{\bf p}_{1}|^{2} -
\mu_{1}^{2}|^{- 1} \times \nonumber \\
|\widetilde{(f_{1}^{0})}_{m_{1}^{(k + 1)}\dot{m}_{1}^{(k + 1)}} (+
0,\cdot) ({\bf p}_{k + 1})| \prod_{s\, =\, 1}^{k} \sum_{n_{2}\, =\,
1,...,4l_{2} + 2} \sum_{\dot{n}_{2} \, =\, 1,...,2\dot{l}_{2} + 1}
\sum_{\nu_{1}, \nu_{2} \, =\, 0}^{3} \sum_{l_{12} \, \in \, 1/2 {\bf
Z}_{+}} \nonumber \\ D_{m_{1}^{(s)}\dot{m}_{1}^{(s)}, m_{1}^{(s +
1)}\dot{m}_{1}^{(s + 1)}}^{1,\nu_{1}} D_{m_{2}\dot{m}_{2},
n_{2}\dot{n}_{2}}^{2,\nu_{2}} D_{\nu_{1} \nu_{2}}^{3,l_{12}}
(2\pi)^{- 8} ((|{\bf p}_{s} - {\bf
p}_{s + 1}|^{2} + \max \lambda_{j}^{2} + 3)^{2} \times \nonumber \\
(1 + (p_{s}^{0} - p_{s + 1}^{0})^{2})^{- 1} \left| \int d^{3}{\bf
y}_{2} (f_{2}^{0})_{m_{2}\dot{m}_{2}} (+ 0,{\bf y}_{2}) \right|^{-
1} |\widetilde{(f_{2}^{0})}_{n_{2}\dot{n}_{2}} (+ 0,\cdot) ({\bf
p}_{s} - {\bf p}_{s + 1})|.
\end{eqnarray}
The Cauchy inequality implies
\begin{equation}
\label{2.84} |{\bf p}_{1}|^{2} \leq \left( \sum_{s\, =\, 1}^{k}
|{\bf p}_{s} - {\bf p}_{s + 1}| + |{\bf p}_{k + 1}| \right)^{2} \leq
(k + 1)\left( \sum_{s\, =\, 1}^{k} |{\bf p}_{s} - {\bf p}_{s +
1}|^{2} + |{\bf p}_{k + 1}|^{2} \right).
\end{equation}
In view of the inequalities (\ref{2.80}), (\ref{2.84}) all integrals
in the right - hand side of the inequality (\ref{2.83}) exist if the
functions $\widetilde{(f_{k}^{0})}_{m_{k}\dot{m}_{k}} (+ 0,\cdot)
({\bf p}_{k})$, $k = 1,2$, are rapidly decreasing at the infinity.
The series (\ref{2.73}) is absolutely convergent. It is the solution
of the equation (\ref{2.71}).

Let us consider the interaction equations for three particles. Let
$\tau$ be a permutation of the numbers $1,2,3$. Let us construct the
interaction equation for $\tau (1)$ particle
\begin{eqnarray}
\label{2.85} \sum_{n_{s} \, =\, 1,...,4l_{s} + 2,\, \, s\, =\,
1,2,3} \sum_{\dot{n}_{s} \, =\, 1,...,2\dot{l}_{s} + 1,\, \, s\, =\,
1,2,3} \int d^{4}x_{\tau (2)} d^{4}x_{\tau (3)} \nonumber
\\ \{ \prod_{s\, =\, 1}^{3} \left( \sum_{\nu \, = \,
0}^{3} (\gamma_{l_{s},\dot{l}_{s}}^{\nu} (A))_{m_{s}\dot{m}_{s},
n_{s}\dot{n}_{s}} \left( - i\frac{\partial}{\partial
x_{s}^{\nu}}\right) + (\beta_{l_{s},\dot{l}_{s}}
(\mu_{s}^{2}))_{m_{s}\dot{m}_{s}, n_{s}\dot{n}_{s}} \right)
(f_{s})_{n_{s}\dot{n}_{s}} (x_{s}) \nonumber \\
- i\prod_{s\, =\, 1}^{3} (\gamma_{l_{s},\dot{l}_{s}}^{0}
(A))_{m_{s}\dot{m}_{s}, n_{s}\dot{n}_{s}} \delta
(x_{s}^{0})(f_{s})_{n_{s}\dot{n}_{s}} (+ 0,{\bf x}_{s})  +
\sum_{j,k\, =\, 2,3,\, j\,
\neq \, k} \sum_{\nu_{1}, \nu_{j} \, =\, 0}^{3} \nonumber \\
A_{\nu_{1} \nu_{j}}^{(\tau (1) \tau (j))} (x_{\tau (1)} - x_{\tau
(j)}) \left( \prod_{s\, =\, 1,j} (\gamma_{l_{\tau (s)},\dot{l}_{\tau
(s)}}^{\nu_{s}} (A))_{m_{\tau (s)}\dot{m}_{\tau (s)}, n_{\tau
(s)}\dot{n}_{\tau (s)}}  \right) \times \nonumber \\
\left( \sum_{\nu \, = \, 0}^{3} (\gamma_{l_{t},\dot{l}_{t}}^{\nu}
(A))_{m_{t}\dot{m}_{t}, n_{t}\dot{n}_{t}} \left( -
i\frac{\partial}{\partial x_{t}^{\nu}}\right) +
(\beta_{l_{t},\dot{l}_{t}} (\mu_{t}^{2}))_{m_{t}\dot{m}_{t},
n_{t}\dot{n}_{t}} \right)_{t = \tau (k)} \prod_{s\, =\, 1}^{3}
(f_{s})_{n_{s}\dot{n}_{s}} (x_{s}) + \nonumber \\
\sum_{\nu_{1}, \nu_{2}, \nu_{3} \, = \, 0}^{3} A_{\nu_{1} \nu_{2}
\nu_{3}}^{(\tau (1))} (x_{\tau (1)} - x_{\tau (2)}, x_{\tau (1)} -
x_{\tau (3)}) \prod_{s\, =\, 1}^{3}
(\gamma_{l_{s},\dot{l}_{s}}^{\nu_{s}} (A))_{m_{s}\dot{m}_{s},
n_{s}\dot{n}_{s}} (f_{s})_{n_{s}\dot{n}_{s}} (x_{s}) \} = 0
\end{eqnarray}
where the potentials $A_{\nu_{1} \nu_{2}}^{(\tau (1) \tau (2))} (x)$
have the form (\ref{2.65}) and the potentials $A_{\nu_{1} \nu_{2}
\nu_{3}}^{(\tau (1))} (x_{1},x_{2})$ satisfy the covariance relation
\begin{equation}
\label{2.86} A_{\nu_{1} \nu_{2} \nu_{3}}^{(\tau (1))} \left(
\sum_{\mu_{s} \, =\, 0}^{3} \Lambda_{\mu_{s}}^{\lambda_{s}} (A^{-
1})x_{s}^{\mu_{s}}, s = 1,2 \right) = \sum_{\mu_{1}, \mu_{2},
\mu_{3} \, =\, 0}^{3} A_{\mu_{1} \mu_{2} \mu_{3}}^{(\tau (1))}
(x_{1},x_{2})\prod_{s\, =\, 1}^{3} \Lambda_{\nu_{s}}^{\mu_{s}} (A)
\end{equation}
for any matrix $A \in SL(2,{\bf C})$. The equation (\ref{2.85})
transforms like three equations (\ref{2.32}).

For any numbers $m,n \in {\bf Z}_{+}$ we define the generalized
Clebsch - Gordan coefficient
\begin{eqnarray}
\label{2.87} C(l_{1},...,l_{m + 2};l_{m + 3},...,l_{m + n +
4};j_{1},...,j_{m + n + 1};
  m_{1},...,m_{m + 2};m_{m + 3},...,m_{m + n + 4}) = \nonumber \\
  \sum_{k_{s}\, \, =\, \, - j_{s}, - j_{s} + 1,...,j_{s} - 1,j_{s},\, \, s\, \, =\, \, 1,...,m + n + 1}
  C(l_{1},l_{2},j_{1};m_{1},m_{2},k_{1}) \times \nonumber \\
  \left( \prod_{s\, =\, 1}^{m} C(j_{s},l_{s + 2},j_{s + 1};k_{s},m_{s + 2},k_{s + 1}) \right)
  \left( \prod_{s\, =\, m + 1}^{m + n} C(j_{s + 1},l_{s + 2},j_{s};k_{s + 1},m_{s + 2},k_{s}) \right)
  \times \nonumber \\
  C(l_{m + n + 4},l_{m + n + 3},j_{m + n + 1};m_{m + n + 4},m_{m + n + 3},k_{m + n + 1})
\end{eqnarray}
where the half - integers $l_{1},...,l_{m + n + 4},j_{1},...,j_{m +
n + 1} \in 1/2{\bf Z}_{+}$ and $m_{i} = - l_{i}, - l_{i} +
1,...,l_{i} - 1,l_{i}, i = 1,...,m + n + 4$. The definition
(\ref{2.87}) and the relations (\ref{2.50}), (\ref{2.51}) imply for
any matrix $A \in SL(2,{\bf C})$
\begin{eqnarray}
\label{2.88} \sum_{n_{s}\, \, =\, \, - l_{s}, - l_{s} + 1,...,l_{s}
- 1,l_{s},\, \, s\, \, =\, \, 1,...,m + 2}
  \left( \prod_{i\, =\, 1}^{m + 2} t_{m_{i},n_{i}}^{l_{i}}(A)\right) \times
  \nonumber \\
C(l_{1},...,l_{m + 2};l_{m + 3},...,l_{m + n + 4};j_{1},...,j_{m + n + 1};
  n_{1},...,n_{m + 2};m_{m + 3},...,m_{m + n + 4}) = \nonumber \\
\sum_{n_{s}\, \, =\, \, - l_{s}, - l_{s} + 1,...,l_{s} - 1,l_{s},\, \, s\, \, =\, \,
m + 3,...,m + n + 4}
  \left( \prod_{i\, =\, m + 3}^{m + n + 4} t_{n_{i},m_{i}}^{l_{i}}(A)\right) \times
  \nonumber \\
C(l_{1},...,l_{m + 2};l_{m + 3},...,l_{m + n + 4};j_{1},...,j_{m + n + 1};
  m_{1},...,m_{m + 2};n_{m + 3},...,n_{m + n + 4}),
\end{eqnarray}
\begin{eqnarray}
\label{2.89} \sum_{n_{s}\, \, =\, \, - l_{s}, - l_{s} + 1,...,l_{s}
- 1,l_{s},\, \, s\, \, =\, \, 1,...,m + 2}
  \left( \prod_{i\, =\, 1}^{m + 2} t_{n_{i},m_{i}}^{l_{i}}(A)\right) \times
  \nonumber \\
C(l_{1},...,l_{m + 2};l_{m + 3},...,l_{m + n + 4};j_{1},...,j_{m + n + 1};
  n_{1},...,n_{m + 2};m_{m + 3},...,m_{m + n + 4}) = \nonumber \\
\sum_{n_{s}\, \, =\, \, - l_{s}, - l_{s} + 1,...,l_{s} - 1,l_{s},\, \, s\, \, =\, \,
m + 3,...,m + n + 4}
  \left( \prod_{i\, =\, m + 3}^{m + n + 4} t_{m_{i},n_{i}}^{l_{i}}(A)\right) \times
  \nonumber \\
C(l_{1},...,l_{m + 2};l_{m + 3},...,l_{m + n + 4};j_{1},...,j_{m + n + 1};
  m_{1},...,m_{m + 2};n_{m + 3},...,n_{m + n + 4}).
\end{eqnarray}
We define the potential
\begin{eqnarray}
\label{2.90} A_{\nu_{1} \nu_{2} \nu_{3}}^{(\tau (1))} (x_{1},x_{2})
= \sum_{l_{12},l_{13},j_{1},j_{2}\, \in \, 1/2{\bf Z}_{+}}
\sum_{m_{1s},\dot{m}_{1s} \, =\, - l_{1s}, - l_{1s} +
1,...,l_{1s} - 1,l_{1s},\, s\, =\, 2,3}  \nonumber \\
\sum_{m_{s},\dot{m}_{s} \, =\, - \frac{1}{2}, \frac{1}{2}, \, s\,
 =\, 1,2,3} C(\frac{1}{2}, \frac{1}{2}, \frac{1}{2};
l_{12},l_{13};j_{1},j_{2};m_{1},m_{2},m_{3};m_{12},m_{13}) \times
 \nonumber \\ C(\frac{1}{2}, \frac{1}{2}, \frac{1}{2};
l_{12},l_{13};j_{1},j_{2};\dot{m}_{1}, \dot{m}_{2}, \dot{m}_{3};
\dot{m}_{12}, \dot{m}_{13}) \left( \prod_{s\, =\, 1}^{3}
t_{m_{s}\dot{m}_{s}}^{\frac{1}{2}} (\sigma^{2} \sigma^{\nu_{s}}
\sigma^{2}) \right) \times \nonumber \\ \int d\lambda_{1} \cdots
d\lambda_{[l_{12} + \frac{3}{2}]} d\kappa_{1} \cdots
d\kappa_{[l_{13} + \frac{3}{2}]} K_{\tau
(1);l_{12},l_{13},j_{1},j_{2}} (\lambda_{1}, ...,\lambda_{[l_{12} +
\frac{3}{2}]}; \kappa_{1}, ...,\kappa_{[l_{13} + \frac{3}{2}]})
\times \nonumber \\ t_{m_{12}\dot{m}_{12}}^{l_{12}}
(\tilde{\partial}_{x_{1}}) e_{\lambda_{1}^{2}, ...,\lambda_{[l_{12}
+ \frac{3}{2}]}^{2}} (x_{1})t_{m_{13}\dot{m}_{13}}^{l_{13}}
(\tilde{\partial}_{x_{2}}) e_{\kappa_{1}^{2}, ...,\kappa_{[l_{13} +
\frac{3}{2}]}^{2}} (x_{2})
\end{eqnarray}
where $[a]$ is the integral part of a real number $a$. In view of
the relations (\ref{2.18}), (\ref{2.37}), (\ref{2.52}),
(\ref{2.53}), (\ref{2.88}), (\ref{2.89}) the potentials (\ref{2.90})
satisfy the covariance relation (\ref{2.86}). We suppose that the
distributions $ K_{\tau (1);l_{12},l_{13},j_{1},j_{2}} (\lambda_{1},
...,\lambda_{[l_{12} + \frac{3}{2}]}; \kappa_{1},
...,\kappa_{[l_{13} + \frac{3}{2}]}) $ have the compact supports and
are not zero for the finite number of the values $l_{12},l_{13}$
only. The definition (\ref{2.87}) and the triangle condition imply
that the generalized Clebsch - Gordan coefficients  are not zero for
the finite number of the values of half - integers $j_{1},j_{2}$
only.

Let $\tau (1)$ and $\tau (2)$ particles interact by means of the
zero mass particles
\begin{eqnarray}
\label{2.91} K_{\tau (1);l_{12},l_{13},j_{1},j_{2}} (\lambda_{1},
...,\lambda_{[l_{12} + \frac{3}{2}]}; \kappa_{1},
...,\kappa_{[l_{13} + \frac{3}{2}]}) = \nonumber \\ K_{\tau
(1);l_{12},l_{13},j_{1},j_{2}} (\kappa_{1}, ...,\kappa_{[l_{13} +
\frac{3}{2}]}) \prod_{s\, =\, 1}^{[l_{12} + \frac{3}{2}]} \delta
(\lambda_{s}).
\end{eqnarray}
The degree of the homogeneous polynomial (\ref{2.5}) is equal to
$2l$. In view of the relations (\ref{2.34}), (\ref{2.36}) the
distribution
\begin{equation}
\label{2.92} t_{m_{12}\dot{m}_{12}}^{l_{12}} (\tilde{\partial}_{x})
e_{0, ...,0} (x)
\end{equation}
with $[l_{12} + \frac{3}{2}]$ zeros has a support in the boundary of
the upper light cone. For $l_{12} = 0$ the integer $[l_{12} +
\frac{3}{2}] = 1$ and the distribution (\ref{2.92}) coincides with
the first distribution (\ref{2.34}). If the integer $2l_{12}
> 1$ is even, then the integer $2[l_{12} + \frac{3}{2}] - 4 =
2l_{12} - 3$. If the integer $2l_{12}$ is odd, then the integer
$2[l_{12} + \frac{3}{2}] - 4 = 2l_{12} - 1$. In view of the
relations (\ref{2.34}), (\ref{2.36}) the distribution (\ref{2.92})
with $[l_{12} + \frac{3}{2}] + 1$ zeros has a support in all closed
upper light cone. If the half - integer $l_{12} = 0$, then the
integer $[l_{12} + \frac{3}{2}] + 1 = 2$ and the second distribution
(\ref{2.34}) has the support in all closed upper light cone. If the
integer $2l_{12} > 1$ is even, then the integer $2[l_{12} +
\frac{3}{2}] - 2 = 2l_{12}$. If the integer $2l_{12}$ is odd, then
the integer $2[l_{12} + \frac{3}{2}] - 2 = 2l_{12} + 1$.

\end{document}